# Quantization of the electromagnetic fields from single atomic or molecular radiators

Valerică Raicu
Department of Physics and Astronomy, University of Wisconsin-Milwaukee, Milwaukee, WI 53211, USA
Email: vraicu@uwm.edu

**Abstract.** A theoretical framework is introduced for expressing electromagnetic (EM) potentials and fields of single atomic or molecular emitters modeled as oscillating dipoles, which follows a recently developed method to solve inhomogeneous wave equations for time-dependent distributions of charge. This framework is first used to evaluate the physical implications of the simplifying assumptions made in the standard approach to the quantization of EM fields and the impact of such assumptions on the predicted energy and momentum quantization. Then, the exact relationships between the potentials and the oscillating (transition) dipole properties derived under the present framework are used to quantize EM fields radiated by single emitters and to achieve agreement with the classical dipole radiation pattern, while maintaining the quantum mechanical description of electromagnetic radiation in terms of the probability distribution of quantum modes. Contributions of the present analysis to the understanding of photon emission from excited atoms or molecules stimulated by light or vacuum field fluctuations are highlighted, and possible experimental tests and practical applications are proposed.

## 1. Introduction

The standard approach to the quantization of the electromagnetic (EM) field, originally introduced by Dirac[1] following a general quantization protocol developed by Heisenberg and others [2,3], starts by expanding the vector potential into a series of plane waves with arbitrary amplitudes, and setting the scalar potential equal to zero, based on a certain implementation of the Coulomb gauge and the assumption that a free field is decoupled from charges and currents that generated it [4-6]. Within this scheme, it may be shown that the EM field is equivalent to a set of harmonic oscillators that are quantized by replacing the amplitudes with creation and annihilation operators obeying quantum mechanical commutation relations. This approach has been essential for understanding absorption and emission of light and led to numerous practical applications, especially in the fields of lasers and photonics [6-8].

However, the introduction of single-molecule fluorescence imaging techniques [9,10] and techniques whereby the relative orientation of transition dipoles within molecular complexes needs to be known [11-13] has made it necessary to relate photon emission statistics to the orientation of the emitting dipoles, particularly the polar angle. Although it may be tempting to assume that the polar angle dependence of photon emission is somehow incorporated into the creation and annihilation operators, the expressions of those operators determined from the Heisenberg equations of motion for the quantum harmonic oscillator [14] have not revealed any such dependence. Therefore, when the necessity to incorporate angular dependence of the radiation emitted either spontaneously or through stimulated emission is recognized [15-19], the problem is usually treated classically rather than quantum mechanically, prompting the need for introduction of a quantum mechanical framework that could handle arbitrarily oriented dipoles. Related to this point is also a recent discussion on stimulated emission [18-20], which is well understood in the context of ensembles of dipole emitters such as gain media in lasers [7], but it needs to be carefully considered in the case of single fluorescent molecule experiments.

In this report, we build on the recent introduction of a method of solving the wave equations for potentials of arbitrary, and evolving, distributions of charge [21] to derive (in Section 2) exact expressions for electromagnetic potentials and fields generated by a single emitter modeled as an oscillating dipole. (The dipole could be thought of as an oscillating superposition of excited and ground state wavefunctions as described in Chapter 3 of the book by Sargent, Scully, and Lamb [7], and it is often used as an excellent model for optically excited atoms or molecules [6,22].) Our derived expressions for potentials and fields contain as their particular cases those traditionally used for quantization of the EM field [4,14], with the notable difference that herein the plane wave amplitudes are actually well defined (as nested integrals of the charge and current distributions over the real and reciprocal space, and time).

This theoretical framework is used (in Section 3) as an ideal testbed for assessing the difficulties posed by the standard quantization framework when applied to single emitters, especially as the specific use of the Coulomb gauge in that framework constrains the direction of the wave vectors $\vec{k}$ of the Fourier components of the field to be perpendicular to the polarization unit vector $\hat{z}$ of the field (see section 4.3 of Ref. [5]), which, as the present work reveals, is equivalent to constraining the direction of the wave vectors to being perpendicular to the emitting dipole; in other words, any dependence of the emitted photons on the polar angle is excluded, which is at variance with the classical dipole radiation pattern [23,24].

The exact classical forms of the potentials and fields are then used (in Section 4) to quantize the EM fields emitted by single dipole radiators. Based on the results of this approach, we feel compelled to conclude that the use of the Coulomb gauge has already fulfilled its role of providing useful guidelines for quantization of free fields from multi-emitter sources such as gain media in lasers, and that its use in connection with single dipole radiators leads to unnecessary complications that can be entirely avoided by using the present causal framework for determination of potentials and fields from charge distributions[21]. In this new framework, the Hamiltonian and momentum operators acquire an explicit dependence on the angle made by the direction of each field mode with the orientation of the emitting dipole.

The theoretical and practical implications of this analysis, including introduction of Planck's constant, will be discussed in Section 5, while some ideas for experimental testing are outlined in the Conclusions section. These results will likely contribute to a better understanding of the processes accompanying radiation emission from excited atoms and molecules in the presence and absence of vacuum fluctuations or stimulating light [14], and a more rigorous interpretation of the results of single molecular dipole imaging [16,17] and quantum mechanics experiments with single photons [8]. In addition, it will facilitate analysis of Förster Resonance Energy Transfer (FRET) experiments without making simplifying assumptions (such as cylindrical-averaging [25]) regarding the orientation of the transitions dipoles within a donor-acceptor complex, or performing costly computer simulations to incorporate instantaneous orientations within the theoretical models [12].

## 2. Scalar and vector potentials for distributions of charge

### 2.1. The k-forms of scalar and vector potentials for an arbitrary distribution of charge

Let us consider an electrical charge distribution (see Fig. 1) that starts forming at $\tau = 0$ at position $\overrightarrow{r'}$ and whose density is $\rho\left(\overrightarrow{r'},\tau\right)\Theta_0(\tau)$ and the corresponding current density is $\vec{j}\left(\overrightarrow{r'},t\right)\Theta_0(t)$, where the step function $\Theta_0(t)$ is equal to zero for $\tau \leq 0$ and 1 for $\tau > 0$. Following the results of a recent publication [21], the scalar potential generated by this distribution at $P_f(\vec{r})$ and satisfies the boundary conditions $\phi(\vec{r},t) = \dot{\phi}(\vec{r},t) = 0$ for $\vec{R} \equiv \vec{r} - \overrightarrow{r'} \rightarrow \infty$ may be written as

$$\phi(\vec{r},t) = \frac{c}{8\pi^3\varepsilon_{\mathrm{v}}}\int_{-\infty}^{t} d\tau \iiint_{-\infty}^{\infty} d^3r'\rho\left(\overrightarrow{r'},\tau\right)\Theta_0(\tau) \iiint_{-\infty}^{\infty} d^3k\,\frac{1}{k}\sin[ck(t-\tau)]\cos\left[\vec{k}\cdot\left(\vec{r}-\overrightarrow{r'}\right)\right], \quad (1)$$

while the vector potentials satisfying the boundary conditions $\vec{A}(\vec{r},t)=\dot{\vec{A}}(\vec{r},t)=0$ for $\vec{R}\equiv\vec{r}-\vec{r'}\to\infty$ may be written as

$$\vec{A}(\vec{r},t)=\frac{c\mu_{\mathrm{v}}}{8\pi^{3}}\int_{-\infty}^{t}d\tau\iiint_{-\infty}^{\infty}d^{3}r'\vec{j}\left(\vec{r'},\tau\right)\Theta_{0}(\tau)\iiint_{-\infty}^{\infty}d^{3}k\frac{1}{k}\sin[ck(t-\tau)]\cos\left[\vec{k}\cdot\left(\vec{r}-\vec{r'}\right)\right]. \qquad (2)$$

These expressions may be rearranged further as (see Appendix A):

$$\phi(\vec{r},t)=\frac{c}{8\pi^{3}\varepsilon_{\mathrm{v}}}\iiint_{-\infty}^{\infty}d^{3}k\frac{1}{k}\int_{-\infty}^{t}d\tau\iiint_{-\infty}^{\infty}d^{3}r'\rho\left(\vec{r'},\tau\right)\Theta_{0}(\tau)\sin\left[kc(t-\tau)-\vec{k}\cdot\left(\vec{r}-\vec{r'}\right)\right], \qquad (3)$$

$$\vec{A}(\vec{r},t)=\frac{c\mu_{\mathrm{v}}}{8\pi^{3}}\iiint_{-\infty}^{\infty}d^{3}k\frac{1}{k}\int_{-\infty}^{t}d\tau\iiint_{-\infty}^{\infty}d^{3}r'\vec{j}\left(\vec{r'},\tau\right)\Theta_{0}(\tau)\sin\left[kc(t-\tau)-\vec{k}\cdot\left(\vec{r}-\vec{r'}\right)\right]. \qquad (4)$$

We will find it useful latter on to expand the sine functions in Eqns. (3) and (4) into exponentials using Euler's formula as

$$\phi(\vec{r},t)=i\frac{c}{16\pi^{3}\varepsilon_{\mathrm{v}}}\iiint_{-\infty}^{\infty}d^{3}k\frac{1}{k}\int_{-\infty}^{t}d\tau\Theta_{0}(\tau)\iiint_{-\infty}^{\infty}d^{3}r'\rho\left(\vec{r'},\tau\right)e^{i\vec{k}\cdot\left(\vec{r}-\vec{r'}\right)-ikc(t-\tau)}-$$

$$i\frac{c}{16\ ^{3}\varepsilon_{\mathrm{v}}}\iiint_{-\infty}^{\infty}d^{3}k\frac{1}{k}\int_{-\infty}^{t}d\tau\Theta_{0}(\tau)\iiint_{-\infty}^{\infty}d^{3}r'\rho\left(\vec{r'},\tau\right)e^{-i\vec{k}\cdot\left(\vec{r}-\vec{r'}\right)+ikc(t-\tau)}, \qquad (5)$$

$$\vec{A}(\vec{r},t)=i\frac{c\mu_{\mathrm{v}}}{16\pi^{3}}\iiint_{-\infty}^{\infty}d^{3}k\frac{1}{k}\int_{-\infty}^{t}d\tau\Theta_{0}(\tau)\iiint_{-\infty}^{\infty}d^{3}r'\vec{j}\left(\vec{r'},\tau\right)e^{i\vec{k}\cdot\left(\vec{r}-\vec{r'}\right)-ikc(t-\tau)}-$$

$$i\frac{c\mu_{\mathrm{v}}}{16\ ^{3}}\iiint_{-\infty}^{\infty}d^{3}k\frac{1}{k}\int_{-\infty}^{t}d\tau\Theta_{0}(\tau)\iiint_{-\infty}^{\infty}d^{3}r'\vec{j}\left(\vec{r'},\tau\right)e^{-i\vec{k}\cdot\left(\vec{r}-\vec{r'}\right)+ikc(t-\tau)}. \qquad (6)$$

Changing the order of integration between $r'$ and $k$, switching to spherical coordinates in $\vec{k}$ by replacing $\vec{k}\cdot\left(\vec{r}-\vec{r'}\right)$ with $k\left|\vec{r}-\vec{r'}\right|\cos\theta$, integrating with respect to the polar angles, and using well-known properties of Dirac's delta function, we recover expressions for the potentials similar to those introduced previously (see Appendix B):

$$\phi(\vec{r},t)=\frac{1}{4\pi\varepsilon_{\mathrm{v}}}\iiint_{-\infty}^{\infty}d^{3}r'\int_{-\infty}^{t}d\tau\Theta_{0}(\tau)\frac{\rho\left(\vec{r'},\tau\right)}{\left|\vec{r}-\vec{r'}\right|}\delta\left[\tau-\left(t-\frac{\left|\vec{r}-\vec{r'}\right|}{c}\right)\right]-$$

$$\frac{1}{4\pi\varepsilon_{\mathrm{v}}}\iiint_{-\infty}^{\infty}d^{3}r'\int_{-\infty}^{t}d\tau\Theta_{0}(\tau)\frac{\rho\left(\vec{r'},\tau\right)}{\left|\vec{r}-\vec{r'}\right|}\delta\left[\tau-\left(t+\frac{\left|\vec{r}-\vec{r'}\right|}{c}\right)\right], \qquad (7)$$

$$\vec{A}(\vec{r},t)=\frac{\mu_{\mathrm{v}}}{4\pi}\iiint_{-\infty}^{\infty}d^{3}r'\int_{-\infty}^{t}d\tau\Theta_{0}(\tau)\frac{\vec{j}\left(\vec{r'},\tau\right)}{\left|\vec{r}-\vec{r'}\right|}\delta\left[\tau-\left(t-\frac{\left|\vec{r}-\vec{r'}\right|}{c}\right)\right]-\frac{\mu_{\mathrm{v}}}{4\pi}\iiint_{-\infty}^{\infty}d^{3}r'\int_{-\infty}^{t}d\tau\Theta_{0}(\tau)\frac{\vec{j}\left(\vec{r'},\tau\right)}{\left|\vec{r}-\vec{r'}\right|}\delta\left[\tau-\right.$$

$$\left.\left(t+\frac{\left|\vec{r}-\vec{r'}\right|}{c}\right)\right]. \qquad (8)$$

These expressions include not only the usual retarded terms but also advanced terms that only act locally (since the second integrals are zero everywhere except in the vicinity of $\vec{R} \equiv \vec{r} - \overrightarrow{r'} = 0$) to cancel out the well-known (and undesired) singularities in $\phi$ and $\vec{A}$ and, thereby, in the electric and magnetic fields [21].

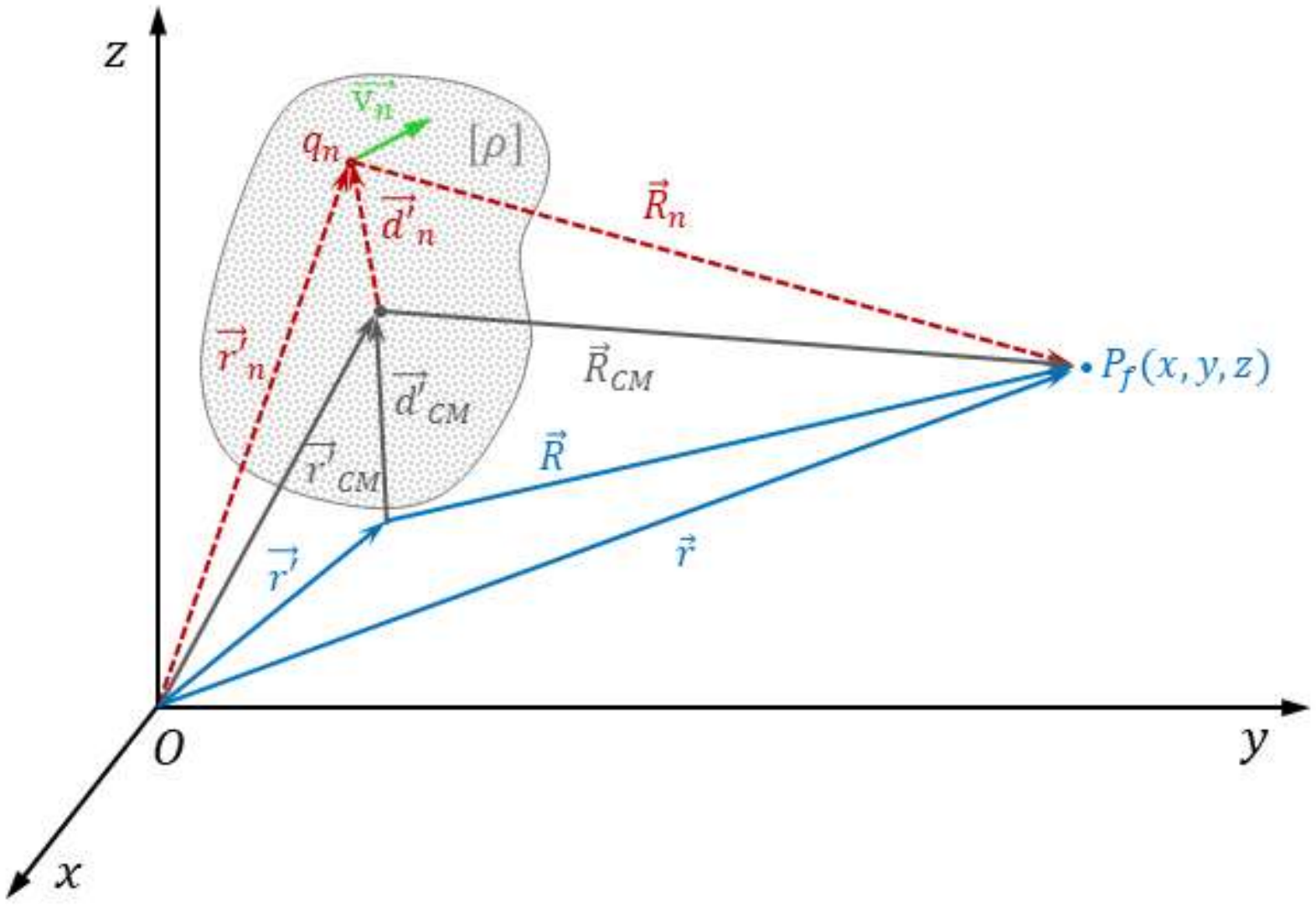

**Figure 1.** Illustration of the geometry used to calculate the potentials and fields (at the field point, $P_f$) generated by a distribution of charge comprised of point charges. The following relations between the different position vectors hold true: $\vec{R} \equiv \vec{r} - \overrightarrow{r'}$ and $\overrightarrow{r'}_n(\tau) = \overrightarrow{r'} + \overrightarrow{d'}_{CM}(\tau) + \overrightarrow{d'}_n(\tau)$.

Alternatively, let us consider a distribution of charge consisting of point charges whose positions within the distribution are defined by delta functions. The charge density for such a distribution is

$$\rho\left(\overrightarrow{r'}, \tau\right) = \sum_n q_n \delta^3\left[\overrightarrow{r'} - \overrightarrow{r'}_n(\tau)\right], \tag{9}$$

while the associated current density is

$$\vec{j}\left(\overrightarrow{r'}, \tau\right) = \sum_n q_n \frac{d\overrightarrow{r'_n}}{dt'} \delta^3\left[\overrightarrow{r'} - \overrightarrow{r'}_n(\tau)\right] = \sum_n q_n \overrightarrow{v_n}(\tau) \delta^3\left[\overrightarrow{r'} - \overrightarrow{r'}_n(\tau)\right]. \tag{10}$$

Upon insertion into Eqns. (5) and (6) and using the sifting property of Dirac's delta function, these densities give

$$\phi(\vec{r}, t) = i \frac{c}{16\pi^3 \varepsilon_v} \iiint_{-\infty}^{\infty} d^3k \frac{1}{k} e^{-ikct} \int_{-\infty}^{t} d\tau \Theta_0(\tau) \sum_n q_n e^{i\vec{k}\cdot\left[\vec{r} - \overrightarrow{r'}_n(\tau)\right] + ikc\tau} -$$

$$i \frac{c}{16\pi^3 \varepsilon_v} \iiint_{-\infty}^{\infty} d^3k \frac{1}{k} e^{ikct} \int_{-\infty}^{t} d\tau \Theta_0(\tau) \sum_n q_n e^{-i\vec{k}\cdot\left[\vec{r} - \overrightarrow{r'}_n(\tau)\right] - ikc\tau}, \tag{11}$$

$$\vec{A}(\vec{r},t) = i\frac{c\mu_{\mathrm{v}}}{16\pi^3}\iiint_{-\infty}^{\infty} d^3k\,\frac{1}{k}\,e^{-ikct}\int_{-\infty}^{t} d\tau\Theta_0(\tau)\sum_n q_n\overrightarrow{\mathrm{v}_n}(\tau)e^{i\vec{k}\cdot\left[\vec{r}-\overrightarrow{r'}_n(\tau)\right]+ik} \quad -$$

$$i\frac{c\mu_{\mathrm{v}}}{16\pi^3}\iiint_{-\infty}^{\infty} d^3k\,\frac{1}{k}\,e^{ikct}\int_{-\infty}^{t} d\tau\Theta_0(\tau)\sum_n q_n\overrightarrow{\mathrm{v}_n}(\tau)e^{-i\vec{k}\cdot\left[\vec{r}-\overrightarrow{r'}_n(\tau)\right]-ikc\tau}. \quad (12)$$

If, using the symbols and relations shown in Figure 1, we write the distance between the "field point" $P_f$ (at position $\vec{r}$ and time *t*) and the "charge point" (at $\overrightarrow{r'}$ and $\tau$) as

$$\vec{r}-\overrightarrow{r'}_n(\tau) = \vec{r}-\overrightarrow{r'}-\overrightarrow{d'}_{CM}(\tau)-\overrightarrow{d'}_n(\tau) \equiv \vec{R}-\overrightarrow{d'}_{CM}(\tau)-\overrightarrow{d'}_n(\tau), \quad (13)$$

where $\overrightarrow{d'}_{CM}(\tau)$ is the distance from the (retarded) point indicated by $\overrightarrow{r'}$ to the center of mass of the atom indicated by $\overrightarrow{r'}_{CM}(\tau)$, $\overrightarrow{d'_n}(\tau)$ is the distance from the center of mass to the charge *n*, and $\vec{R}$ is the distance between the field point and the instantaneous position of the charge (i.e., a time-independent property of space), Eqns. (11) and (12) may be rearranged as

$$\phi(\vec{r},t) = i\frac{c}{16\pi^3\varepsilon_{\mathrm{v}}}\iiint_{-\infty}^{\infty} d^3k\,\frac{1}{k}\,e^{i\vec{k}\cdot\vec{R}-ikct}\int_{-\infty}^{t} d\tau\Theta_0(\tau)e^{ikc\tau-i\vec{k}\cdot\overrightarrow{d'}_{CM}(\tau)}\sum_n q_n e^{-i\vec{k}\cdot\overrightarrow{d'}_n(\tau)} \; -$$

$$i\frac{c}{16\pi^3\varepsilon_{\mathrm{v}}}\iiint_{-\infty}^{\infty} d^3k\,\frac{1}{k}\,e^{-i\vec{k}\cdot\vec{R}+ikct}\int_{-\infty}^{t} d\tau\Theta_0(\tau)e^{-ikc\tau+i\vec{k}\cdot\overrightarrow{d'}_{CM}(\tau)}\sum_n q_n e^{i\vec{k}\cdot\overrightarrow{d'}_n(\tau)}, \quad (14)$$

$$\vec{A}(\vec{r},t) = i\frac{c\mu_{\mathrm{v}}}{16\pi^3}\iiint_{-\infty}^{\infty} d^3k\,\frac{1}{k}\,e^{i\vec{k}\cdot\vec{R}-ikct}\int_{-\infty}^{t} d\tau\Theta_0(\tau)e^{ikc\tau-i\vec{k}\cdot\overrightarrow{d'}_{CM}(\tau)}\sum_n q_n\overrightarrow{\mathrm{v}_n}(\tau)e^{-i\vec{k}\cdot\overrightarrow{d'}_n(\tau)} \; -$$

$$i\frac{c\mu_{\mathrm{v}}}{16\ {}^{3}}\iiint_{-\infty}^{\infty} d^3k\,\frac{1}{k}\,e^{-i\vec{k}\cdot\vec{R}+ikct}\int_{-\infty}^{t} d\tau\Theta_0(\tau)e^{-ikc\tau+i\vec{k}\cdot\overrightarrow{d'}_{CM}(\tau)}\sum_n q_n\overrightarrow{\mathrm{v}_n}(\tau)e^{i\vec{k}\cdot\overrightarrow{d'}_n(\tau)}. \quad (15)$$

The last two equations may be used to compute the potentials for complex distributions of charge and thereby their electric and magnetic fields.

## 2.2. Scalar and vector potentials of oscillating dipoles

While the forms of the potentials given by Eqns. (7) and (8) provide insights into the electrodynamics of distributions of charge [21], their reciprocal-space (or k-space) representations given by Eqns. (14) and (15) are more amenable to the study of charge coupling to radiation. This treatment circumvents some vexing issues posed by introduction of (point) dipoles in classical electrodynamics [26]. Next, we will use the last set of expressions to write the k-forms of the scalar and vector potential for the case of a single oscillating dipole, defined in a general way as a pair of point charges separated by a variable distance. Since excited atoms and molecules may be well described by oscillating dipoles[7,22], these will allow us to quantize the EM field emitted by single atoms or molecules.

Under the oscillating dipole model (see Fig. 2) and using the summation indices for position and velocity from Eqns. (14) and (15), one of the charges (e.g., $q_0 = +e$, where $e$ is the absolute charge of the electron), which represents the nucleus, is placed at the instantaneous position $\overrightarrow{d'}_0(\tau) = -z'_+(\tau)\hat{z}$ and moves with velocity $\overrightarrow{\mathrm{v}_0}(\tau) = -\mathrm{v}_+(\tau)\hat{z}$ relative to the center of mass, *CM*, of the atom. The second charge ($q_1 = -e$), representing the center of mass of the electron cloud, is placed at the instantaneous position $\overrightarrow{d'}_1(\tau) = z'_-(\tau)\hat{z}$ and moves with velocity $\overrightarrow{\mathrm{v}_1}(\tau) = \mathrm{v}_-(\tau)\hat{z}$. For simplicity, we assume, without loss of generality, that the motions of the charges within the dipoles occur along the *z* axis. In addition, the subscript $CM$, which stands for the center of mass of the entire distribution of charge (i.e., the electron-nucleus pair), is now replaced by $d$, to more suggestively refer to the dipole.

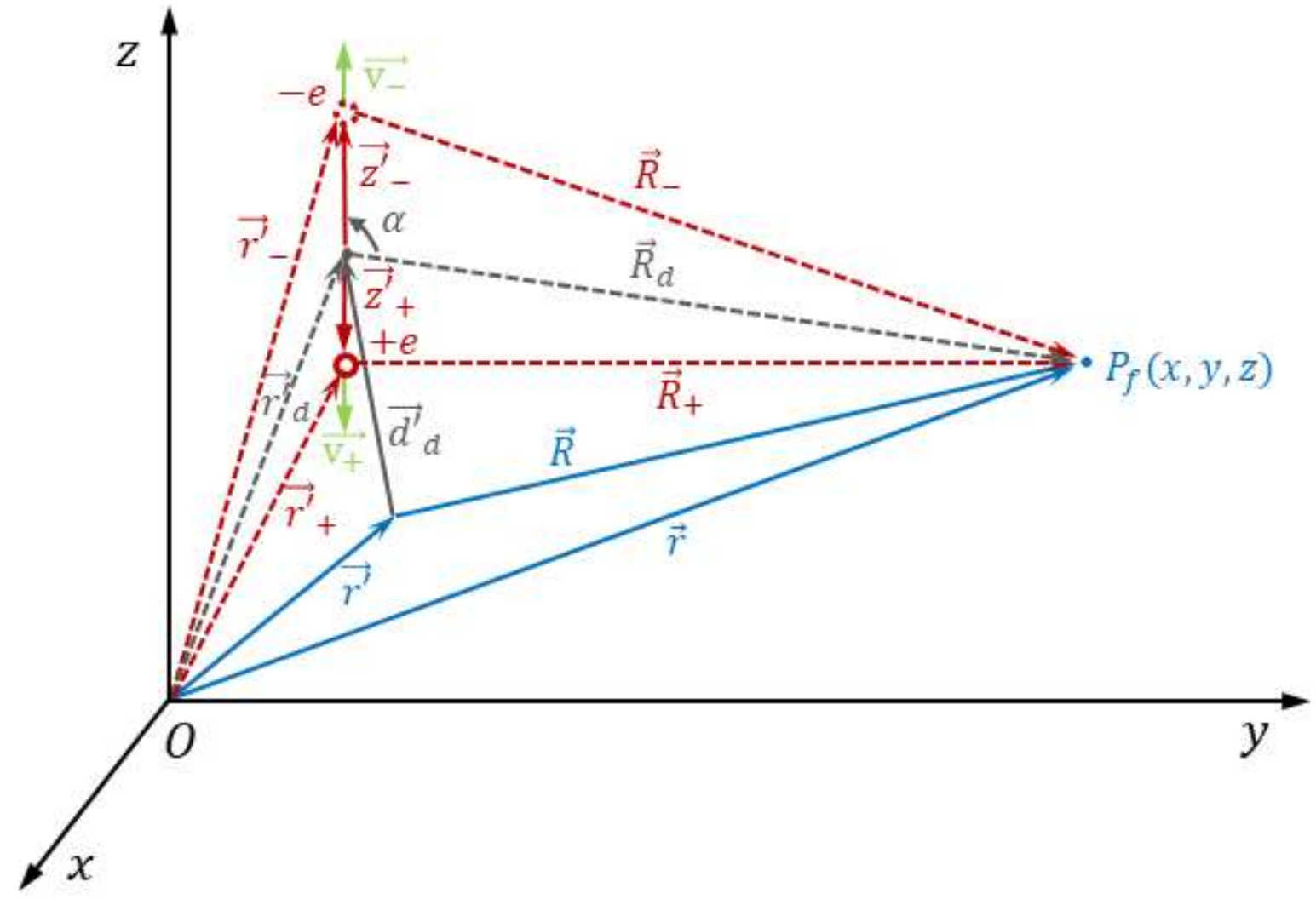

**Figure 2.** Geometry used to calculate the potentials and fields (at the field point, $P_f$) generated by an oscillating dipole comprised of a positive and a negative point charge. For the negative charge, we will use $\cos\alpha = \hat{z}\cdot\hat{R}_d$ when writing the cosine law for the top triangle, while for the positive charge, we use $\cos(\pi-\alpha) = -\hat{z}\cdot\hat{R}_d$ when writing the cosine law for the bottom triangle. The following relations between the different position vectors hold true: $\overrightarrow{r'_\pm}(\tau) = \overrightarrow{r_d'} + \overrightarrow{z'_\pm}(\tau)$, $\overrightarrow{r_d'} = \vec{r'} + \overrightarrow{z_d'}$, and $\overrightarrow{R_\pm}(\tau) = \vec{r} - \overrightarrow{r'_\pm}(\tau) = \overrightarrow{R_d} - \overrightarrow{z_\pm'}(\tau) = \vec{R} - \overrightarrow{z_d'} - \overrightarrow{z_\pm'}(\tau)$. Note that $z'_+$ and $z'_-$ are not shown to scale, since actually $z'_+ \ll z'_-$, due to the large difference between the nucleus and electron mass.

For the assumed dipole, Eqns. (14) and (15) give the exact expressions

$$\phi(\vec{r},t) = \iiint_{-\infty}^{\infty} d^3k \left[\mathcal{F}\left(\vec{k},t\right) e^{i\vec{k}\cdot\vec{R}-ikct} + \mathcal{F}^*\left(\vec{k},t\right) e^{-i\vec{k}\cdot\vec{R}+ikct}\right], \qquad (16)$$

where

$$\mathcal{F}\left(\vec{k},t\right) = i\,\frac{ec}{16\pi^3\varepsilon_{\mathrm{v}}}\frac{1}{k}\int_{-\infty}^{t} d\tau\Theta_0(\tau)e^{ikc\tau-\;\vec{}\cdot\overrightarrow{d_d'}(\tau)}\left[e^{-i\vec{k}\cdot\overrightarrow{z_+'}(\tau)} - e^{-i\vec{k}\cdot\overrightarrow{z_-'}(\tau)}\right], \tag{17}$$

and

$$\mathcal{F}^*\left(\vec{k},t\right) = -i\,\frac{ec}{16\pi^3\varepsilon_{\mathrm{v}}}\frac{1}{k}\int_{-\infty}^{t} d\tau\Theta_0(\tau)e^{-ikc\tau+i\vec{k}\cdot\overrightarrow{d_d'}(\tau)}\left[e^{i\vec{k}\cdot\overrightarrow{z_+'}(\tau)} - e^{i\vec{k}\cdot\overrightarrow{z_-'}(\tau)}\right], \tag{18}$$

and

$$\vec{A}(\vec{r},t) = \iiint_{-\infty}^{\infty} d^3k\left[\vec{\mathcal{A}}\left(\vec{k},t\right)e^{i\vec{k}\cdot\vec{R}-ikct} + \overrightarrow{\mathcal{A}^*}\left(\vec{k},t\right)e^{-i\vec{k}\cdot\vec{R}+ikct}\right], \tag{19}$$

where

$$\vec{\mathcal{A}}\left(\vec{k},t\right) = i\,\frac{ec\mu_{\mathrm{v}}}{16\pi^3}\frac{1}{k}\int_{-\infty}^{t} d\tau\Theta_0(\tau)e^{ikc\tau-\;\vec{}\cdot\overrightarrow{d_d'}(\tau)}\left[\overrightarrow{\mathrm{v}_+}(\tau)e^{-i\vec{k}\cdot\overrightarrow{z_+'}(\tau)} - \overrightarrow{\mathrm{v}_-}(\tau)e^{-i\vec{k}\cdot\overrightarrow{z_-'}(\tau)}\right], \tag{20}$$

and

$$\overrightarrow{\mathcal{A}^*}\left(\vec{k},t\right) = -i\,\frac{ec\mu_{\mathrm{v}}}{16\pi^3}\frac{1}{k}\int_{-\infty}^{t} d\tau\Theta_0(\tau)e^{-ikc\tau+i\vec{k}\cdot\overrightarrow{d_d'}(\tau)}\left[\overrightarrow{\mathrm{v}_+}(\tau)e^{i\vec{k}\cdot\overrightarrow{z_+'}(\tau)} - \overrightarrow{\mathrm{v}_-}(\tau)e^{i\vec{k}\cdot\overrightarrow{z_-'}(\tau)}\right]. \tag{21}$$

### 2.3. Separation of the scalar potential into position- and velocity-dependent terms

While in the previous sub-section the scalar potential has been explicitly defined via $\mathcal{F}\left(\vec{k},t\right)$ and $\mathcal{F}^*\left(\vec{k},t\right)$ given by Eqns. (17) and (18), respectively, it is possible to separate the potential into terms that are velocity dependent – and are therefore $\vec{\mathcal{A}}\left(\vec{k},t\right)$ and $\overrightarrow{\mathcal{A}^*}\left(\vec{k},t\right)$ dependent –, and terms that are dependent on distances only, as briefly illustrated next and described in detail in Appendix C.

Taking the dot product between $\vec{k}$ and $\vec{\mathcal{A}}\left(\vec{k},t\right)$ given by Eqn. (20), noticing that $-i\hat{k}\cdot\left[\overrightarrow{\mathrm{v}_+}(\tau)e^{-i\vec{k}\cdot\overrightarrow{z_+'}(\tau)} - \overrightarrow{\mathrm{v}_-}(\tau)e^{-i\vec{k}\cdot\overrightarrow{z_-'}(\tau)}\right] = \frac{d}{d\tau}\left[e^{-i\vec{k}\cdot\overrightarrow{z_+'}(\tau)} - e^{-i\vec{k}\cdot\overrightarrow{z_-'}(\tau)}\right]$, integrating by parts, and using well-known properties of the delta function, we obtain

$$\vec{k}\cdot\vec{\mathcal{A}}\left(\vec{k},t\right) = -\frac{ec\mu_{\mathrm{v}}}{16\pi^3}\frac{1}{k}e^{ikct-i\vec{k}\cdot\overrightarrow{d_d'}(t)}\left[e^{-i\vec{k}\cdot\overrightarrow{z_+'}(t)} - e^{-i\vec{k}\cdot\overrightarrow{z_-'}(t)}\right] +$$
$$i\,\frac{e}{16\pi^3\varepsilon_{\mathrm{v}}}\int_{-\infty}^{t} d\tau\Theta_0(\tau)e^{ikc\tau-i\vec{k}\cdot\overrightarrow{d_d'}(\tau)}\left[e^{-i\vec{k}\cdot\overrightarrow{z_+'}(\tau)} - e^{-i\vec{k}\cdot\overrightarrow{z_-'}(\tau)}\right] - i\,\frac{e}{16\pi^3\varepsilon_{\mathrm{v}}}\int_{-\infty}^{t} d\tau\Theta_0(\tau)\left[\hat{k}\cdot\right.$$
$$\left.\overrightarrow{\mathrm{v}_d}(\tau)\right]e^{ikc\tau-i\vec{k}\cdot\overrightarrow{d_d'}(\tau)}\left[e^{-i\vec{k}\cdot\overrightarrow{z_+'}(\tau)} - e^{-i\vec{k}\cdot\overrightarrow{z_-'}(\tau)}\right]. \tag{22}$$

Substituting Eqn. (17) into the last expression and rearranging of terms we obtain

$$\mathcal{F}(\vec{k},t) = c\hat{k}\cdot\vec{\mathcal{A}}(\vec{k},t) + \frac{e}{16\pi^3\varepsilon_\mathrm{v}}\frac{1}{k^2}e^{ikct-i\vec{k}\cdot\overrightarrow{d'_d}(t)}\left[e^{-i\vec{k}\cdot\overrightarrow{z'_+}(t)} - e^{-i\vec{k}\cdot\overrightarrow{z'_-}(t)}\right] + c\hat{k}\cdot\vec{\mathcal{V}}(\vec{k},t), \tag{23}$$

with

$$\mathcal{V}(\vec{k},t) = i\frac{ec\mu_\mathrm{v}}{16\pi^3}\frac{1}{k}\int_{-\infty}^{t} d\tau\Theta_0(\tau)\overrightarrow{\mathrm{v}_d}e^{ikc\tau-i\vec{k}\cdot\overrightarrow{d'_d}(\tau)}\left[e^{-i\vec{k}\cdot\overrightarrow{z'_+}(\tau)} - e^{-i\vec{k}\cdot\overrightarrow{z'_-}(\tau)}\right]. \tag{24}$$

Similarly, taking the dot product of $\vec{k}$ with $\overrightarrow{\mathcal{A}^*}(\vec{k},t)$ in Eqn. (21), integrating by parts, and using Eqn. (18), we obtain

$$\mathcal{F}^*(\vec{k},t) = c\hat{k}\cdot\vec{\mathcal{A}}^*(\vec{k},t) + \frac{e}{16\pi^3\varepsilon_\mathrm{v}}\frac{1}{k^2}e^{-ikct+i\vec{k}\cdot\overrightarrow{d'_d}(t)}\left[e^{i\vec{k}\cdot\overrightarrow{z'_+}(t)} - e^{i\vec{k}\cdot\overrightarrow{z'_-}(t)}\right] + c\hat{k}\cdot\vec{\mathcal{V}}^*(\vec{k},t), \tag{25}$$

with

$$\vec{\mathcal{V}}^*(\vec{k},t) = -i\frac{ec\mu_\mathrm{v}}{16\pi^3}\frac{1}{k}\int_{-\infty}^{t} d\tau\Theta_0(\tau)\overrightarrow{\mathrm{v}_d}(\tau)e^{-ikc\tau+i\vec{k}\cdot\overrightarrow{d'_d}(\tau)}\left[e^{i\vec{k}\cdot\overrightarrow{z'_+}(\tau)} - e^{i\vec{k}\cdot\overrightarrow{z'_-}(\tau)}\right]. \tag{26}$$

Upon insertion of Equations (23) and (25) into (16), we obtain

$$\phi(\vec{r},t) = \iiint_{-\infty}^{\infty} d^3k\; c\hat{k}\cdot\left[\vec{\mathcal{A}}(\vec{k},t)e^{i\vec{k}\cdot\vec{R}-ikct} + \overrightarrow{\mathcal{A}^*}(\vec{k},t)e^{-i\vec{k}\cdot\vec{R}+ikct}\right] + \frac{e}{4\pi\varepsilon_\mathrm{v}R_+(t)} - \frac{e}{4\pi\varepsilon_\mathrm{v}R_-(t)} +$$
$$\iiint_{-\infty}^{\infty} d^3k\; c\hat{k}\cdot\left[\vec{\mathcal{V}}(\vec{k},t)e^{i\vec{k}\cdot\vec{R}-ikct} + \vec{\mathcal{V}}^*(\vec{k},t)e^{-i\vec{k}\cdot\vec{R}+ikct}\right], \tag{27}$$

where $R_-(t)$ and $R_+(t)$ are the distances between each of the two charges comprising the dipole and the field position at the present time *t*. To be meticulous, it should be remarked that $R_-(t)$ and $R_+(t)$ may only be known approximately to the observer for cases where retardation is negligible (i.e., for short distances to the field point). Regardless, the two terms incorporating $R_-(t)$ and $R_+(t)$ store the potential energy of the excited dipole in the surrounding space and is confined to comparatively short distances. This energy is dissipated through the two terms containing the nested integrals (over both $k$-space and time, via $\vec{\mathcal{A}}$, $\overrightarrow{\mathcal{A}^*}$, $\vec{\mathcal{V}}$, and $\vec{\mathcal{V}}^*$), which, together with the vector potential may be used to compute the part of the electric field that propagates to large distances (see below). If the dipole has emitted the entire stored energy by the time of measurement, $t > 0$, then $R_-(t)$ and $R_+(t)$ become exactly equal to $R_d$ (i.e., the dipole vanishes by that time) and the two terms cancel each other out. This implies that the EM field may now be regarded as a "free field."

## 2.4. Workplan for the remainder of the paper

In the next section, we will show that the well-known results of the electromagnetic field quantization may be derived within the present theoretical framework by using the strong small-dipole approximation in Eqns. (16)-(21). This will allow us to take advantage of the expressions of the complex amplitudes $\vec{\mathcal{A}}(\vec{k},t)$ and $\overrightarrow{\mathcal{A}^*}(\vec{k},t)$ that the present theoretical framework provides, in order to reach a better understanding of the approximations and tradeoffs implicit in the standard approach to quantization [5]. Those constraints will then be relaxed in section 4, to derive more general expressions that explicitly take into account the orientation of the emitting dipoles to gain physical insights into stimulated and spontaneous emission, which have led and will likely continue to lead to many practical applications.

## 3. Assessing the standard approach to fields quantization within the present framework

When the quantization of the EM field was first introduced by Dirac [1], the exact forms of the expansion coefficients given by Eqns. (17), (18), (20) and (21) had not been known. Instead, the vector potential has been expanded into a sum of plane waves with constant coefficients, and the scalar potential has been assumed to be equal to zero, which is usually justified by application of the Coulomb gauge and the assumption that a free field is decoupled from charges and currents that produced it, i.e., charge density and longitudinal current densities, may be set to zero [4,5,14]. To emulate those results, we use herein the *strong small-dipole approximation*, whereby $\overrightarrow{z'_+}(\tau)$ in Eqns. (17)-(19) is negligible when compared to $\vec{R}$, so that the complex amplitudes $\mathcal{F}(\vec{k},t)$ and $\mathcal{F}^*(\vec{k},t)$ vanish. We obtain for a dipole whose center of mass is fixed at the position $z'_d$ (see Appendix D):

$$\phi(\vec{r},t) = 0, \tag{28}$$

$$\vec{A}(\vec{r},t) = \iiint_{-\infty}^{\infty} d^3k\, \hat{z} \left[\mathcal{A}(\vec{k},t) e^{i\vec{k}\cdot\vec{R}-ik} \quad + \mathcal{A}^*(\vec{k},t) e^{-i\vec{k}\cdot\vec{R}+ikct}\right], \tag{29}$$

$$\vec{E}(\vec{r},t) = i \iiint_{-\infty}^{\infty} d^3k\, kc\, \hat{z} \left[\mathcal{A}(\vec{k},t) e^{i\vec{k}\cdot\vec{R}-ikct} - \mathcal{A}^*(\vec{k},t) e^{-i\vec{k}\cdot\vec{R}+ikct}\right], \tag{30}$$

$$\vec{B}(\vec{r},t) = i \iiint_{-\infty}^{\infty} d^3k\, (\vec{k}\times\hat{z}) \left[\mathcal{A}(\vec{k},t) e^{i\vec{k}\cdot\vec{R}-ikct} - \mathcal{A}^*(\vec{k},t) e^{-i\vec{k}\cdot\vec{R}+ikct}\right], \tag{31}$$

where we chose the direction of the dipole to coincide with the z-axis (see Fig. 2) and therefore wrote: $\vec{\mathcal{A}}(\vec{k},t) = \hat{z}\mathcal{A}(\vec{k},t)$ and $\overrightarrow{\mathcal{A}^*}(\vec{k},t) = \hat{z}\mathcal{A}^*(\vec{k},t)$. The last equations are formally identical to the well-known expressions for the vector potential and fields used in the quantization of the 'free' EM fields, except that here the amplitudes $\mathcal{A}(k,t)$ and $\mathcal{A}^*(k,t)$ are known [see Eqns. (20) and (21)], and we use only one vector to denote the polarization of the EM field: $\hat{z}$. Optical polarization could of course be decomposed into two orthogonal vectors; in that case, strictly speaking, we would have to consider quadrupoles instead of dipoles, an unnecessary complication that we will avoid in this work.

We will find it convenient to use the notations

$$\alpha(\vec{k},t) = \mathcal{A}(\vec{k},t)e^{-ikc} \ , \tag{32}$$

$$\alpha^*(\vec{k},t) = \mathcal{A}^*(\vec{k},t)e^{ikct}, \tag{33}$$

to write Eqns. (30) and (31) as

$$\vec{E}(\vec{r},t) = i \iiint_{-\infty}^{\infty} d^3k \ kc \ \hat{z} \left[\alpha(\vec{k},t)e^{i\vec{k}\cdot\vec{R}} - \alpha^*(\vec{k},t)e^{-i\vec{k}\cdot\vec{R}}\right], \tag{34}$$

$$\vec{B}(\vec{r},t) = i \iiint_{-\infty}^{\infty} d^3k\left(\vec{k}\times\hat{z}\right)\left[\alpha(\vec{k},t)e^{i\vec{k}\cdot\vec{R}} - \alpha^*(\vec{k},t)e^{-i\vec{k}\cdot\vec{R}}\right]. \tag{35}$$

The total electromagnetic energy integrated over the entire space centered around the distribution of charge that emitted it is

$$H = \frac{\varepsilon_{\mathrm{v}}}{2}\iiint_{-\infty}^{\infty} d^3R\left(\vec{E}\cdot\vec{E}\right) + \frac{1}{2\mu_{\mathrm{v}}}\iiint_{-\infty}^{\infty} d^3R\left(\vec{B}\cdot\vec{B}\right) =$$
$$-\frac{\varepsilon_{\mathrm{v}}}{2}c^2\iiint_{-\infty}^{\infty} d^3R\iiint_{-\infty}^{\infty} d^3k_1 d^3k \ k_1 k\left[\alpha(\vec{k}_1,t)e^{i\overrightarrow{k_1}\cdot\vec{R}} - \alpha^*(\vec{k}_1,t)e^{-i\overrightarrow{k_1}\cdot\vec{R}}\right]\left[\alpha(\vec{k},t)e^{i\vec{k}\cdot\vec{R}} - \right.$$
$$\left.\alpha^*(\vec{k},t)e^{-i\vec{k}\cdot\vec{R}}\right] - \frac{\varepsilon_{\mathrm{v}}}{2}c^2\iiint_{-\infty}^{\infty} d^3R\iiint_{-\infty}^{\infty} d^3k_1 d^3k\left\{\left(\vec{k}_1\times\hat{z}\right)\left[\alpha(\vec{k}_1,t)e^{i\overrightarrow{k_1}\cdot\vec{R}} - \alpha^*(\vec{k}_1,t)e^{-i\overrightarrow{k_1}\cdot\vec{R}}\right]\right\}\cdot$$
$$\left\{\left(\vec{k}\times\hat{z}\right)\left[\alpha(\vec{k},t)e^{i\vec{k}\cdot\vec{R}} - \alpha^*(\vec{k},t)e^{-i\vec{k}\cdot\vec{R}}\right]\right\}. \tag{36}$$

Using the vector identity $\left(\vec{a}\times\vec{b}\right)\cdot\left(\vec{c}\times\vec{d}\right) = (\vec{a}\cdot\vec{c})\left(\vec{b}\cdot\vec{d}\right) - \left(\vec{a}\cdot\vec{d}\right)\left(\vec{b}\cdot\vec{c}\right)$ for the second set of integrals, changing the order of integration, and recognizing the resulting integrals over volume as Dirac delta functions,

$$\iiint_{-\infty}^{\infty} d^3R e^{\pm i\left(\overrightarrow{k_1}+\vec{k}\right)\cdot\vec{R}} = 8\pi^3\delta^3\left(\overrightarrow{k_1}+\vec{k}\right), \tag{37}$$

$$\iiint_{-\infty}^{\infty} d^3R e^{\pm i\left(\overrightarrow{k_1}-\vec{k}\right)\cdot\vec{R}} = 8\pi^3\delta^3\left(\overrightarrow{k_1}-\vec{k}\right). \tag{38}$$

we obtain

$$H = 8\pi^3\varepsilon_{\mathrm{v}}\iiint_{-\infty}^{\infty} d^3k \ \omega_k{}^2\left[\alpha(\vec{k},t)\alpha^*(\vec{k},t) + \alpha^*(\vec{k},t)\alpha(\vec{k},t)\right] - 4\pi^3\varepsilon_{\mathrm{v}}c^2\iiint_{-\infty}^{\infty} d^3k\left(\vec{k}\cdot\right.$$
$$\left.\hat{z}\right)^2\left[\alpha(\vec{k},t)\alpha(\vec{k},t) + \alpha(\vec{k},t)\alpha^*(\vec{k},t) + \alpha^*(\vec{k},t)\alpha(\vec{k},t) + \alpha^*(\vec{k},t)\alpha^*(\vec{k},t)\right]. \tag{39}$$

where $\omega_k = kc$ is the angular frequency of the mode *k*.

The standard approach to the quantization of the EM field is to choose at this point the direction of $\vec{k}$ to be perpendicular to the polarization unit vector $\hat{z}$ for each mode *k* [5], which gives

$$\vec{k}\cdot\hat{z}\alpha(\vec{k},t) = \vec{k}\cdot\hat{z}\alpha^*(\vec{k},t) = 0. \tag{40}$$

Under this condition, the terms comprising the dot products containing $\vec{k} \cdot \hat{z}$ vanish, while $\hat{z} \cdot \hat{z} = 1$. Note that the present framework – wherein $\vec{\mathcal{A}}$ and $\overrightarrow{\mathcal{A}^*}$ have been derived from first principles and are defined in terms of the velocities of the charges comprising the dipoles – makes it evident that this orthonormality assumption considers only modes whose wave vectors are perpendicular to the dipole moment. This assumption is not suitable to use for randomly oriented single emitters, which should follow the classical dipole radiation pattern [16,23,24]. However, for the time being, we will adopt the orthonormality condition, which we will dispose of when we take up the more general approach introduced in the next section. Using condition (40), the total energy carried away by the electromagnetic radiation emitted by a vibrating dipole is

$$H = 8\pi^3 \varepsilon_{\mathrm{v}} \iiint_{-\infty}^{\infty} d^3k\, {\omega_k}^2 \left[\alpha(\vec{k},t)\alpha^*(\vec{k},t) + \alpha^*(\vec{k},t)\alpha(\vec{k},t)\right] =$$

$$8\pi^3 \varepsilon_{\mathrm{v}} \iiint_{-\infty}^{\infty} d^3k\, {\omega_k}^2 \left[\mathcal{A}(\vec{k},t)\mathcal{A}^*(\vec{k},t) + \mathcal{A}^*(\vec{k},t)\mathcal{A}(\vec{k},t)\right], \tag{41}$$

where we chose not to use the fact that the two products in the square bracket are obviously commutative in the present (classical) case.

If we next express the two complex amplitudes $\mathcal{A}(\vec{k},t)$ and $\mathcal{A}^*(\vec{k},t)$ as functions of two real valued vectors $q(t)$ and $p(t)$, i.e.,

$$\alpha(\vec{k},t) = \frac{\mathrm{V}^{1/2}}{(256\pi^6 \varepsilon_{\mathrm{v}} m)^{1/2} \omega_k} \left[m\omega_k q(\vec{k},t) + ip(\vec{k},t)\right], \tag{42}$$

$$\alpha^*(\vec{k},t) = \frac{\mathrm{V}^{1/2}}{(256\pi^6 \varepsilon_{\mathrm{v}} m)^{1/2} \omega_k} \left[m\omega_k q(\vec{k},t) - ip(\vec{k},t)\right], \tag{43}$$

Eqn. (41) becomes

$$H = \frac{\mathrm{V}}{8\pi^3} \iiint_{-\infty}^{\infty} d^3k \left(\frac{p^2}{2m} + \frac{1}{2} m {\omega_k}^2 q^2\right). \tag{44}$$

Since this equation is similar to the Hamiltonian of the harmonic oscillator, the physical interpretation of notations (42) and (43) is that the field mode $k$ is equivalent to a harmonic oscillator with oscillation frequency $\omega_k$, position $q(k,t)$, momentum $p(k,t)$, and mass $m$ [5,14]. This implies that the solutions to the system of equations (42) and (43),

$$q(\vec{k},t) = \sqrt{\frac{256\pi^6 \varepsilon_{\mathrm{v}}}{m\mathrm{V}}} \left[\alpha(\vec{k},t) + \alpha^*(\vec{k},t)\right], \tag{45}$$

$$p(\vec{k},t) = -i\omega_k \sqrt{\frac{256\pi^6 \varepsilon_{\mathrm{v}} m}{\mathrm{V}}} \left[\alpha(\vec{k},t) - \alpha^*(\vec{k},t)\right], \tag{46}$$

must be canonically conjugate variables, i.e., they obey Hamilton's equations for the harmonic oscillator:

$$\dot{q}(\vec{k},t) = \frac{1}{m}p(\vec{k},t), \tag{47}$$

$$\dot{p}(\vec{k},t) = -m{\omega_k}^2 q(\vec{k},t). \tag{48}$$

It may be easily verified using Eqns. (20), (21), (32), and (33) that the position and momentum expressions (45) and (46) do indeed obey equations (47) and (48) (Appendix E) for times greater than $t_1$ at which the oscillatory motion of the dipole ceases, i.e., $\overrightarrow{\mathrm{v}_+}(t) = \overrightarrow{\mathrm{v}_-}(t) = 0$. The fields, which continue to propagate after the emitting dipole oscillations have ceased, are similar to those referred to in textbooks as "free fields." As a result of this constraint, the upper limit on the temporal integrals in $\mathcal{A}(\vec{k},t)$ and $\mathcal{A}^*(\vec{k},t)$ is now replaced by $t_1$ and the dependence of these amplitudes on $t$ will be dropped hereafter.

We can now proceed to quantizing the EM field using the equations introduced above as the starting point, and replacing the generalized $p$ and $q$ coordinates in Eqn. (45)-(48) [2,3] by the momentum and position operators such that the quantity given by Eqn. (44) becomes the (Heisenberg) Hamiltonian of the quantum harmonic oscillator. This may be done operationally by replacing the amplitudes $\mathcal{A}(k)$ and $\mathcal{A}^*(k)$ in (32)-(35) by the annihilation, $\boldsymbol{a}(\vec{k})$, and creation, $\boldsymbol{a}^\dagger(\vec{k})$, operators [4,5,14] via

$$\mathcal{A}(\vec{k}) \rightarrow \left(\frac{\hbar\mathrm{V}}{128\pi^6\varepsilon_\mathrm{v}\omega_k}\right)^{1/2} \boldsymbol{a}(\vec{k}) \tag{49}$$

and

$$\mathcal{A}^*(\vec{k}) \rightarrow \left(\frac{\hbar\mathrm{V}}{128\pi^6\varepsilon_\mathrm{v}\omega_k}\right)^{1/2} \boldsymbol{a}^\dagger(\vec{k}), \tag{50}$$

where the operators obey the following commutation relations [4], derived in Appendix F, for convenience:

$$\left[\boldsymbol{a}(\vec{k}), \boldsymbol{a}^\dagger(\vec{k}')\right] = \delta_{\vec{k}\vec{k}'}, \left[\boldsymbol{a}(\vec{k}), \boldsymbol{a}(\vec{k}')\right] = \left[\boldsymbol{a}^\dagger(\vec{k}), \boldsymbol{a}^\dagger(\vec{k}')\right] = 0. \tag{51}$$

The integrals over the k-space are often replaced by sums according to the rules

$$\vec{k} = \frac{2\pi}{L}\left(l_x\hat{k}_x + l_y\hat{k}_y + l_z\hat{k}_z\right) \tag{52}$$

with $l_x, l_y, l_z = 0, \pm1, \pm2, \pm3, \ldots$, and

$$\frac{1}{(2\pi)^3}\iiint_{-\infty}^{\infty} d^3k\,(\quad) \rightarrow \frac{1}{V}\sum_{\vec{k}}(\quad), \tag{53}$$

although it is often advantageous to retain the integral forms, which we will continue doing in this report.

Thus, the classical Hamiltonian in Eqn. (41) may be replaced by the Hamiltonian of the quantum harmonic oscillator [4],

$$\boldsymbol{H} = \frac{1}{2}\frac{\mathrm{V}}{8\pi^3}\iiint_{-\infty}^{\infty} d^3k\, \hbar\omega_k\left[\boldsymbol{a}(\vec{k})\boldsymbol{a}^{\dagger}(\vec{k}) + \boldsymbol{a}^{\dagger}(\vec{k})\boldsymbol{a}(\vec{k})\right] = \mathrm{V}\iiint_{-\infty}^{\infty} d^3k\, \hbar\omega_k\left[\boldsymbol{a}^{\dagger}(\vec{k})\boldsymbol{a}(\vec{k}) + \frac{1}{2}\right], \qquad (54)$$

which obeys the commutation relation (51), and the scalar potential and the EM fields become

$$\vec{\boldsymbol{A}}(\vec{r},t) = \left(\frac{\hbar\mathrm{V}}{128\pi^6\varepsilon_{\mathrm{V}}}\right)^{1/2}\iiint_{-\infty}^{\infty} d^3k\, \hat{z}\omega_k{}^{-1/2}\left[\boldsymbol{a}(\vec{k})e^{i\vec{k}\cdot\vec{R}-ikct} + \boldsymbol{a}^{\dagger}(\vec{k})e^{-i\vec{k}\cdot\vec{R}+ikc}\ \right], \qquad (55)$$

$$\vec{\boldsymbol{E}}(\vec{r},t) = -i\left(\frac{\hbar\mathrm{V}}{128\pi^6\varepsilon_{\mathrm{V}}}\right)^{1/2}\iiint_{-\infty}^{\infty} d^3k\, \hat{z}\omega_k{}^{1/2}\left[\boldsymbol{a}(\vec{k})e^{i\vec{k}\cdot\vec{R}-ikct} - \boldsymbol{a}^{\dagger}(\vec{k})e^{-i\vec{k}\cdot\vec{R}+ikct}\right], \qquad (56)$$

$$\vec{\boldsymbol{B}}(\vec{r},t) = i\left(\frac{\hbar\mathrm{V}}{128\pi^6\varepsilon_{\mathrm{V}}}\right)^{1/2}\iiint_{-\infty}^{\infty} d^3k\, (\vec{k}\times\hat{z})\omega_k{}^{-1/2}\left[\boldsymbol{a}(\vec{k})e^{i\vec{k}\cdot\vec{R}-ikc}\ - \boldsymbol{a}^{\dagger}(\vec{k})e^{-i\vec{k}\cdot\vec{R}+ikct}\right]. \qquad (57)$$

Our derivation above makes it rather clear that the orthonormality relations given by Eqns. (40) should be considered as additional conditions that may not be physically correct. Specifically, the orthonormality relations imply that the emitted electromagnetic waves (or photons) propagate only in directions perpendicular to the emitting dipole (i.e., within a plane perpendicular to the dipole) and therefore deviate from the classical dipole radiation pattern. This will become more obvious looking at the expression for the momentum which we will derive next.

Using Eqns. (26) and (27), the well-known identity $\vec{a}\times(\vec{b}\times\vec{c}) = (\vec{a}\cdot\vec{c})\vec{b} - (\vec{a}\cdot\vec{b})\vec{c}$, the relations $\vec{\alpha}(\vec{k},t) = \hat{z}\alpha(\vec{k},t)$, $\overrightarrow{\alpha^*}(\vec{k},t) = \hat{z}\alpha^*(\vec{k},t)$, $\vec{k}\cdot\hat{z} = 0$ and $\hat{z}\cdot\hat{z} = 1$, the electromagnetic momentum [27] integrated over the entire space may be written successively as

$$\vec{G} = \varepsilon_{\mathrm{V}}\iiint_{-\infty}^{\infty} d^3R\,(\vec{E}\times\vec{B}) = -\varepsilon_{\mathrm{V}}c\iiint_{-\infty}^{\infty} d^3R\iiint_{-\infty}^{\infty} d^3k_1 d^3k\, k_1\left\{\left[\alpha(\vec{k}_1,t)e^{i\vec{k_1}\cdot\vec{R}} - \alpha^*(\vec{k}_1,t)e^{-i\vec{k_1}\cdot\vec{R}}\right]\left[\alpha(\vec{k},t)e^{i\vec{k}\cdot\vec{R}} - \alpha^*(\vec{k},t)e^{-i\vec{k}\cdot\vec{R}}\right]\vec{k}\right\} + \varepsilon_{\mathrm{V}}c\iiint_{-\infty}^{\infty} d^3R\iiint_{-\infty}^{\infty} d^3k_1 d^3k\, k_1(\vec{k}\cdot\hat{z})\left[\alpha(\vec{k}_1,t)e^{i\vec{k_1}\cdot\vec{R}} - \alpha^*(\vec{k}_1,t)e^{-i\vec{k_1}\cdot\vec{R}}\right]\left[\alpha(\vec{k},t)e^{i\vec{k}\cdot\vec{R}} - \alpha^*(\vec{k},t)e^{-i\vec{k}\cdot\vec{R}}\right]. \qquad (58)$$

After performing the multiplications, changing the order of integration, replacing the integrals over volume by the Dirac delta functions defined by Eqns. (37) and (38), and using the sifting property of the delta function and the orthonormality relations (40) (for consistency with the derivation of the Hamiltonian), we obtain

$$\vec{G} = 8\pi^3\varepsilon_{\mathrm{V}}\iiint_{-\infty}^{\infty} d^3k\,\omega_k\vec{k}\left[\alpha(\vec{k},t)\alpha^*(\vec{k},t) + \alpha^*(\vec{k},t)\alpha(\vec{k},t)\right] - 8\pi^3\varepsilon_{\mathrm{V}}\iiint_{-\infty}^{\infty} d^3k\,\omega_k\vec{k}\left[\alpha(\vec{k},t)\vec{\alpha}(\vec{k},t)\right] - 8\pi^3\varepsilon_{\mathrm{V}}\iiint_{-\infty}^{\infty} d^3k\,\omega_k\vec{k}\left[\alpha^*(\vec{k},t)\alpha^*(\vec{k},t)\right], \qquad (59)$$

where $\omega_k = kc$. By quantizing the momentum in its current form, using the scheme introduced above, wherein $\mathcal{A}(\vec{k})$ and $\mathcal{A}^*(\vec{k})$ are replaced by the $\boldsymbol{a}_{\vec{k}}$ and $\boldsymbol{a}_{\vec{k}}^{\dagger}$, respectively, we get

$$\vec{G} = \frac{1}{2}\frac{V}{8\pi^3}\iiint_{-\infty}^{\infty} d^3k\, \hbar\vec{k}\left[\boldsymbol{a}(\vec{k})\boldsymbol{a}^\dagger(\vec{k}) + \boldsymbol{a}^\dagger(\vec{k})\boldsymbol{a}(\vec{k})\right] - \frac{1}{2}\frac{V}{8\pi^3}\iiint_{-\infty}^{\infty} d^3k\, \hbar\vec{k}\left[\boldsymbol{a}(\vec{k})\boldsymbol{a}(\vec{k}) + \boldsymbol{a}^\dagger(\vec{k})\boldsymbol{a}^\dagger(\vec{k})\right]. \quad (60)$$

Using the second set of commutation relations (51), the two terms in the second integral vanish, and Eqn. (60) becomes

$$\vec{G} = \frac{1}{2}\frac{V}{8\pi^3}\iiint_{-\infty}^{\infty} d^3k\, \hbar\vec{k}\left[\boldsymbol{a}(\vec{k})\boldsymbol{a}^\dagger(\vec{k}) + \boldsymbol{a}^\dagger(\vec{k})\boldsymbol{a}(\vec{k})\right] = \frac{V}{8\pi^3}\iiint_{-\infty}^{\infty} d^3k\, \hbar\vec{k}\left[\boldsymbol{a}^\dagger(\vec{k})\boldsymbol{a}(\vec{k}) + \frac{1}{2}\right], \quad (61)$$

where we have also used the first commutation relation in Eqn. (51).

A somewhat subtle point here is that the second integral in Eqn. (59) is only made to disappear through some "quantum magic," i.e., by using the second commutation relation in Eqn. (51). It also becomes apparent now but not entirely surprising that, since the electromagnetic momentum of Eqn. (61) points in an arbitrary direction, so should $\vec{k}$, which contradicts the assumption made in deriving Eqn. (61) that $\vec{k}$ is restricted to directions perpendicular to $\hat{z}$. These inconsistencies seem to indicate that the standard QED's implementation of the Coulomb gauge is questionable when applied to single radiators. We will show in the following section that both of these difficulties are inevitably addressed by making weaker approximations in the derivation of the electric field.

## 4. Quantization of the free EM fields of single dipoles

In the previous section, in neglecting the scalar potential completely, we inadvertently discarded from the final results important velocity-dependent terms incorporated in Eqn. (27), which make contributions to the electric field that are of similar magnitude to those originating from the vector potential and cannot therefore be ignored. We will use Eqn. (27) under the simplifying assumption that the center of mass of the dipole is instantaneously at rest (i.e., $\vec{v_d} = 0$), which removes the last set of integrals from Eqn. (27). In addition, we will ignore the non-propagating part of the field, which is obtained from the second and third terms of the scalar potential (27). (These two terms resembling the electrostatic potential could be also exactly set to zero if, by the time of measurement, *t*, the dipole has emitted the entire electromagnetic energy stored in its surrounding space. In that case, the EM field may be regarded as a "free field," which is also equivalent, loosely speaking, to a statement on quantization of the EM field, since the oscillator necessarily emits a finite amount of energy between the time at which it starts oscillating and the time when it stops. Under these assumptions, and ignoring retardation (i.e., using instantaneous derivatives) as we already have in the previous section, the electric field obtained by summing up the negative gradient of the scalar potential given by Eqn. (27) and the negative temporal derivative of the vector potential of Eqn. (19) is readily obtained as:

$$\vec{E}(\vec{r},t) = -i \iiint_{-\infty}^{\infty} d^3k\, \omega_k \hat{k}\left(\hat{k}\cdot\hat{z}\right) \left[\mathcal{A}\left(\vec{k},t\right)e^{i\vec{k}\cdot\vec{R}-ikct} - \mathcal{A}^*\left(\vec{k},t\right)e^{-i\vec{k}\cdot\vec{R}+ikct}\right] +$$
$$i \iiint_{-\infty}^{\infty} d^3k\, \omega_k \hat{z} \left[\mathcal{A}\left(\vec{k},t\right)e^{i\vec{k}\cdot\vec{R}-ikct} - \mathcal{A}^*\left(\vec{k},t\right)e^{-i\vec{k}\cdot\vec{R}+ikct}\right], \tag{62}$$

where $\omega_k = kc$ and we used again the assumption that the charge oscillations occur along the z axis and therefore $\vec{\mathcal{A}}\left(\vec{k},t\right) = \hat{z}\mathcal{A}\left(\vec{k},t\right)$ and $\overrightarrow{\mathcal{A}^*}\left(\vec{k},t\right) = \hat{z}\mathcal{A}^*\left(\vec{k},t\right)$.

Using the well-known vector identity for the triple vector product, the last expression may be rewritten as

$$\vec{E}(\vec{r},t) = -i \iiint_{-\infty}^{\infty} d^3k\, \omega_k\, \hat{k}\times\left(\hat{k}\times\hat{z}\right) \left[\mathcal{A}\left(\vec{k},t\right)e^{i\vec{k}\cdot\vec{R}-ikct} - \mathcal{A}^*\left(\vec{k},t\right)\, e^{-i\vec{k}\cdot\vec{R}+ikc}\ \right]. \tag{63}$$

At the same time, the magnetic field [obtained as the curl of Eqn. (19)] remains the same as before,

$$\vec{B}(\vec{r},t) = i \iiint_{-\infty}^{\infty} d^3k\left(\vec{k}\times\hat{z}\right) \left[\mathcal{A}\left(\vec{k},t\right)\, e^{i\vec{k}\cdot\vec{R}-ikct} - \mathcal{A}^*\left(\vec{k},t\right)\, e^{-i\vec{k}\cdot\vec{R}+ikct}\right]. \tag{64}$$

After using the re-notations given by Eqns. (34) and (35), in which $\vec{\alpha}(k,t) = \hat{z}\alpha(k,t)$ and $\overrightarrow{\alpha^*}\left(\vec{k},t\right) = \hat{z}\alpha^*\left(\vec{k},t\right)$ as in the previous section, equations (63) and (64) become

$$\vec{E}(\vec{r},t) = -i \iiint_{-\infty}^{\infty} d^3k\, \omega_k\, \hat{k}\times\left(\hat{k}\times\hat{z}\right) \left[\alpha\left(\vec{k},t\right)\, e^{i\vec{k}\cdot\vec{R}} - \alpha^*\left(\vec{k},t\right)\, e^{-i\vec{k}\cdot\vec{R}}\right], \tag{65}$$

$$\vec{B}(\vec{r},t) = i \iiint_{-\infty}^{\infty} d^3k\, k\left(\hat{k}\times\hat{z}\right) \left[\alpha\left(\vec{k},t\right)e^{i\vec{k}\cdot\vec{R}} - \alpha^*\left(\vec{k},t\right)e^{-i\vec{k}\cdot\vec{R}}\right]. \tag{66}$$

With the last two expressions, the total electromagnetic energy integrated over the entire space centered around the dipole may be written as

$$H = \frac{\varepsilon_{\mathrm{v}}}{2} \iiint_{-\infty}^{\infty} d^3R\, \left(\vec{E}\cdot\vec{E}\right) + \frac{1}{2\mu_{\mathrm{v}}} \iiint_{-\infty}^{\infty} d^3R\, \left(\vec{B}\cdot\vec{B}\right) = -\frac{\varepsilon_{\mathrm{v}}}{2} \iiint_{-\infty}^{\infty} d^3R \iiint_{-\infty}^{\infty} d^3k_1 d^3k\, \omega_{k_1}\omega_k \left[\hat{k}_1 \times\right.$$
$$\left.\left(\hat{k}_1\times\hat{z}\right)\right]\cdot\left[\hat{k}\times\left(\hat{k}\times\hat{z}\right)\right] \left[\alpha\left(\vec{k}_1,t\right)e^{i\overrightarrow{k_1}\cdot\vec{R}} - \alpha^*\left(\vec{k}_1,t\right)e^{-i\overrightarrow{k_1}\cdot\vec{R}}\right] \left[\alpha\left(\vec{k},t\right)e^{i\vec{k}\cdot\vec{R}} - \alpha^*\left(\vec{k},t\right)e^{-i\vec{k}\cdot\vec{R}}\right] -$$
$$\frac{\varepsilon_{\mathrm{v}}}{2} \iiint_{-\infty}^{\infty} d^3R \iiint_{-\infty}^{\infty} d^3k_1 d^3k\, \omega_{k_1}\omega_k \left[\left(\hat{k}_1\times\hat{z}\right)\cdot\left(\hat{k}\times\hat{z}\right)\right] \left[\alpha\left(\vec{k}_1,t\right)e^{i\overrightarrow{k_1}\cdot\vec{R}} -\right.$$
$$\left.\alpha^*\left(\vec{k}_1,t\right)e^{-i\overrightarrow{k_1}\cdot\vec{R}}\right] \left[\alpha\left(\vec{k},t\right)e^{i\vec{k}\cdot\vec{R}} - \alpha^*\left(\vec{k},t\right)e^{-i\vec{k}\cdot\vec{R}}\right], \tag{67}$$

and, after using the vector identities $\vec{a}\times\left(\vec{b}\times\vec{c}\right) = (\vec{a}\cdot\vec{c})\vec{b} - \left(\vec{a}\cdot\vec{b}\right)\vec{c}$ and $\left(\vec{a}\times\vec{b}\right)\cdot\left(\vec{c}\times\vec{d}\right) = (\vec{a}\cdot\vec{c})\left(\vec{b}\cdot\vec{d}\right) - \left(\vec{a}\cdot\vec{d}\right)\left(\vec{b}\cdot\vec{c}\right)$ and changing the order of integration,

$$H = -\frac{\varepsilon_{\mathrm{v}}}{2} \iiint_{-\infty}^{\infty} d^3k_1 d^3k\, \omega_{k_1}\omega_k \left\{\left[\left(\hat{k}_1\cdot\hat{z}\right)\hat{k}_1 - \hat{z}\right]\cdot\left[\left(\hat{k}\cdot\hat{z}\right)\hat{k} - \hat{z}\right] + \left(\hat{k}_1\cdot\hat{k}\right) - \left(\hat{k}_1\cdot\hat{z}\right)\left(\hat{k}\cdot\right.\right.$$
$$\left.\left.\hat{z}\right)\right\} \iiint_{-\infty}^{\infty} d^3R \left[\alpha\left(\vec{k}_1,t\right)e^{i\overrightarrow{k_1}\cdot\vec{R}} - \alpha^*\left(\vec{k}_1,t\right)e^{-i\overrightarrow{k_1}\cdot\vec{R}}\right] \left[\alpha\left(\vec{k},t\right)e^{i\vec{k}\cdot\vec{R}} - \alpha^*\left(\vec{k},t\right)e^{-i\vec{k}\cdot\vec{R}}\right]. \tag{68}$$

Performing the multiplications under the spatial integrals and grouping the spatial exponentials, we obtain

$$H = -\frac{\varepsilon_{\mathrm{v}}}{2}\iiint_{-\infty}^{\infty} d^3k_1 d^3k\ \omega_{k_1}\omega_k\{[(\hat{k}_1\cdot\hat{z})\hat{k}_1-\hat{z}]\cdot[(\hat{k}\cdot\hat{z})\hat{k}-\hat{z}]+(\hat{k}_1\cdot\hat{k})-(\hat{k}_1\cdot\hat{z})(\hat{k}\cdot\hat{z})\}\iiint_{-\infty}^{\infty} d^3R\left[\alpha(\vec{k}_1,t)\alpha(\vec{k},t)e^{i(\overrightarrow{k_1}+\vec{k})\cdot\vec{R}}-\alpha(\vec{k}_1,t)\alpha^*(\vec{k},t)e^{i(\overrightarrow{k_1}-\vec{k})\cdot\vec{R}}-\alpha^*(\vec{k}_1,t)\alpha(\vec{k},t)e^{-i(\overrightarrow{k_1}-\vec{k})\cdot\vec{R}}+\alpha^*(\vec{k}_1,t)\alpha^*(\vec{k},t)e^{-i(\overrightarrow{k_1}+\vec{k})\cdot\vec{R}}\right]. \tag{69}$$

By expressing the integrals over volume in terms of the Dirac delta functions as given by Eqns. (37) and (38), and using the sifting property of the delta function, we obtain

$$H = 8\pi^3\varepsilon_{\mathrm{v}}\iiint_{-\infty}^{\infty} d^3k\ {\omega_k}^2\left[1-(\hat{k}\cdot\hat{z})^2\right]\left[\alpha(\vec{k},t)\alpha^*(\vec{k},t)+\alpha^*(\vec{k},t)\alpha(\vec{k},t)\right] = 8\pi^3\varepsilon_{\mathrm{v}}\iiint_{-\infty}^{\infty} d^3k\ {\omega_k}^2\left[1-(\hat{k}\cdot\hat{z})^2\right]\left[\mathcal{A}(\vec{k},t)\mathcal{A}^*(\vec{k},t)+\mathcal{A}^*(\vec{k},t)\mathcal{A}(\vec{k},t)\right], \tag{70}$$

where we also used

$$[(\hat{k}\cdot\hat{z})\hat{k}-\hat{z}]\cdot[(\hat{k}\cdot\hat{z})\hat{k}-\hat{z}]-(\hat{k}\cdot\hat{k})+(\hat{k}\cdot\hat{z})^2 = (\hat{k}\cdot\hat{z})^2-(\hat{k}\cdot\hat{z})^2-(\hat{k}\cdot\hat{z})^2+1-1+(\hat{k}\cdot\hat{z})^2 = 0, \tag{71}$$

and

$$[(\hat{k}\cdot\hat{z})\hat{k}-\hat{z}]\cdot[(\hat{k}\cdot\hat{z})\hat{k}-\hat{z}]+(\hat{k}\cdot\hat{k})-(\hat{k}\cdot\hat{z})^2 = (\hat{k}\cdot\hat{z})^2-(\hat{k}\cdot\hat{z})^2-(\hat{k}\cdot\hat{z})^2+1+1-(\hat{k}\cdot\hat{z})^2 = 2\left[1-(\hat{k}\cdot\hat{z})^2\right]. \tag{72}$$

Equation (70) is valid for any arbitrary angle $\varkappa_{\vec{k}}$ between $\vec{k}$ and $\vec{z}$. Using this angle explicitly, we write $\hat{k}\cdot\hat{z}^2 = \cos\varkappa_{\vec{k}}$, and obtain the expression

$$H = 8\pi^3\varepsilon_{\mathrm{v}}\iiint_{-\infty}^{\infty} d^3k\ {\omega_k}^2\sin^2\varkappa_{\vec{k}}\left[\mathcal{A}(\vec{k},t)\mathcal{A}^*(\vec{k},t)+\mathcal{A}^*(\vec{k},t)\mathcal{A}(\vec{k},t)\right], \tag{73}$$

which predicts that the energy of every mode $k$ depends not only on its wavelength (or frequency) but also on the angle under which the radiation is emitted (or detected).

Next, using Eqns. (56) and (57), the electromagnetic momentum integrated over the entire space may be written successively as

$$\vec{G} = \varepsilon_{\mathrm{v}}\iiint_{-\infty}^{\infty} d^3R\ (\vec{E}\times\vec{B}) = \varepsilon_{\mathrm{v}}\iiint_{-\infty}^{\infty} d^3R\iiint_{-\infty}^{\infty} d^3k_1 d^3k\ \omega_{k_1}k[(\hat{k}_1\cdot\hat{z})\hat{k}_1-\hat{z}]\times(\hat{k}\times\hat{z})\left[\vec{\alpha}(\vec{k}_1,t)e^{i\overrightarrow{k_1}\cdot\vec{R}}-\overrightarrow{\alpha^*}(\vec{k}_1,t)e^{-i\overrightarrow{k_1}\cdot\vec{R}}\right]\left[\alpha(\vec{k},t)e^{i\vec{k}\cdot\vec{R}}-\alpha^*(\vec{k},t)e^{-i\vec{k}\cdot\vec{R}}\right] = \varepsilon_{\mathrm{v}}\iiint_{-\infty}^{\infty} d^3k_1 d^3k\ \omega_{k_1}k\left[(\hat{k}_1\cdot\hat{z})^2\hat{k}-(\hat{k}_1\cdot\hat{z})(\hat{k}_1\cdot\hat{k})\hat{k}_1-\hat{k}+(\hat{k}\cdot\hat{z})\hat{z}\right]\iiint_{-\infty}^{\infty} d^3R\left[\vec{\alpha}(\vec{k}_1,t)e^{i\overrightarrow{k_1}\cdot\vec{R}}-\overrightarrow{\alpha^*}(\vec{k}_1,t)e^{-i\overrightarrow{k_1}\cdot\vec{R}}\right]\left[\alpha(\vec{k},t)e^{i\vec{k}\cdot\vec{R}}-\alpha^*(\vec{k},t)e^{-i\vec{k}\cdot\vec{R}}\right], \tag{74}$$

where we used $\vec{a}\times(\vec{b}\times\vec{c}) = (\vec{a}\cdot\vec{c})\vec{b}-(\vec{a}\cdot\vec{b})\vec{c}$ to convert the vector products of the unit vectors and changed the order of integration. Performing the multiplications under the spatial integral, and combining spatial exponentials, we obtain

$$\vec{G} = \varepsilon_{\mathrm{v}} \iiint_{-\infty}^{\infty} d^3k_1 d^3k\, \omega_{k_1} k \left[\left(\hat{k}_1 \cdot \hat{z}\right)^2 \hat{k} - \left(\hat{k}_1 \cdot \hat{z}\right)\left(\hat{k}_1 \cdot \hat{k}\right)\hat{z} - \hat{k} + \left(\hat{k} \cdot \hat{z}\right)\hat{z}\right] \left[\alpha\left(\vec{k}_1, t\right)\alpha\left(\vec{k}, t\right) \iiint_{-\infty}^{\infty} d^3R\, e^{i\left(\vec{k_1}+\vec{k}\right)\cdot\vec{R}} - \alpha\left(\vec{k}_1, t\right)\alpha^*\left(\vec{k}, t\right) \iiint_{-\infty}^{\infty} d^3R\, e^{i\left(\vec{k_1}-\vec{k}\right)\cdot\vec{R}} - \alpha^*\left(\vec{k}_1, t\right)\alpha\left(\vec{k}, t\right) \iiint_{-\infty}^{\infty} d^3R\, e^{-i\left(\vec{k_1}-\vec{k}\right)\cdot\vec{R}} + \alpha^*\left(\vec{k}_1, t\right)\alpha^*(k, t) \iiint_{-\infty}^{\infty} d^3R\, e^{-i\left(\vec{k_1}+\vec{k}\right)\cdot\vec{R}}\right]. \tag{75}$$

Replacing the integrals over volume by the Dirac delta functions defined by Eqns. (37) and (38) and using the sifting property of the delta function, the last expression becomes

$$\vec{G} = 8\pi^3 \varepsilon_{\mathrm{v}} \iiint_{-\infty}^{\infty} d^3k\, \omega_k \vec{k} \left[1 - \left(\hat{k} \cdot \hat{z}\right)^2\right] \left[\alpha\left(\vec{k}, t\right)\alpha^*\left(\vec{k}, t\right) + \alpha^*\left(\vec{k}, t\right)\alpha\left(\vec{k}, t\right)\right] - 8\pi^3 \varepsilon_{\mathrm{v}} \iiint_{-\infty}^{\infty} d^3k\, \omega_k \vec{k} \left[1 - \left(\hat{k} \cdot \hat{z}\right)^2\right] \left[\alpha\left(-\vec{k}, t\right)\alpha\left(\vec{k}, t\right) + \alpha^*\left(-\vec{k}, t\right)\alpha^*\left(\vec{k}, t\right)\right]. \tag{76}$$

By splitting the second integral into two integrals, we obtain

$$\iiint_{-\infty}^{\infty} d^3k\, \omega_k \vec{k} \left[1 - \left(\hat{k} \cdot \hat{z}\right)^2\right] \left[\alpha\left(-\vec{k}, t\right)\alpha\left(\vec{k}, t\right) + \alpha^*\left(-\vec{k}, t\right)\alpha^*\left(\vec{k}, t\right)\right] = - \iiint_{0}^{-\infty} d^3k\, \omega_k \vec{k} \left[1 - \left(\hat{k} \cdot \hat{z}\right)^2\right] \left[\alpha\left(-\vec{k}, t\right)\alpha\left(\vec{k}, t\right) + \alpha^*\left(-\vec{k}, t\right)\alpha^*\left(\vec{k}, t\right)\right] + \iiint_{0}^{\infty} d^3k\, \omega_k \vec{k} \left[1 - \left(\hat{k} \cdot \hat{z}\right)^2\right] \left[\alpha\left(-\vec{k}, t\right)\alpha\left(\vec{k}, t\right) + \alpha^*\left(-\vec{k}, t\right)\alpha^*\left(\vec{k}, t\right)\right] = - \iiint_{0}^{\infty} d^3k\, \omega_k \vec{k} \left[1 - \left(\hat{k} \cdot \hat{z}\right)^2\right] \left[\alpha\left(\vec{k}, t\right)\alpha\left(-\vec{k}, t\right) + \alpha^*\left(\vec{k}, t\right)\alpha^*\left(-\vec{k}, t\right)\right] + \iiint_{0}^{\infty} d^3k\, \omega_k \vec{k} \left[1 - \left(\hat{k} \cdot \hat{z}\right)^2\right] \left[\alpha\left(-\vec{k}, t\right)\alpha\left(\vec{k}, t\right) + \alpha^*\left(-\vec{k}, t\right)\alpha^*\left(\vec{k}, t\right)\right] = 0, \tag{77}$$

where we also used a change of variable $-\vec{k} \rightarrow \vec{k}$ along the way. Therefore, the second integral in (76) vanishes, giving

$$\vec{G} = 8\pi^3 \varepsilon_{\mathrm{v}} \iiint_{-\infty}^{\infty} d^3k\, \omega_k k\, \hat{k} \left[1 - \left(\hat{k} \cdot \hat{z}\right)^2\right] \left[\alpha\left(\vec{k}, t\right)\alpha^*\left(\vec{k}, t\right) + \alpha^*\left(\vec{k}, t\right)\alpha\left(\vec{k}, t\right)\right], \tag{78}$$

or, using again $\hat{k} \cdot \hat{z} = \cos \varkappa_{\vec{k}}$ and Eqns. (32) and (33) (in which $t$ is replaced by a constant, $t_I$),

$$\vec{G} = 8\pi^3 \varepsilon_{\mathrm{v}} \iiint_{-\infty}^{\infty} d^3k\, \omega_k \vec{k} \sin^2 \varkappa_{\vec{k}} \left[\mathcal{A}\left(\vec{k}\right)\mathcal{A}^*\left(\vec{k}\right) + \mathcal{A}^*\left(\vec{k}\right)\mathcal{A}\left(\vec{k}\right)\right]. \tag{79}$$

Equations (73) and (79) predict that the energy and momentum of mode $k$ depend not only on the wavelength (or frequency) of the mode, but also on the angle under which the radiation is emitted (or detected). This angle ensures that there is no radiation emission along the direction of the dipole moment and in general restores the agreement with the dipole radiation pattern already known from the real-space representation of the field in classical electrodynamics [23,24]. The quantized forms of the potentials [Eqns. (19) and (27), with appropriate approximations], fields [Eqns. (63) and (64)], Hamiltonian [Eqn. (73)], and momentum [Eqns. (79)] are given by the expressions:

$$\phi(\vec{r}, t) = \left(\frac{\hbar \mathrm{V}}{128\pi^6 \varepsilon_{\mathrm{v}}}\right)^{1/2} \iiint_{-\infty}^{\infty} d^3k\, \left(\hat{k} \cdot \hat{z}\right) c\, \omega_k^{-1/2} \left[\boldsymbol{a}\left(\vec{k}\right) e^{i\vec{k}\cdot\vec{R} - ikc} + \boldsymbol{a}^\dagger\left(\vec{k}\right) e^{-i\vec{k}\cdot\vec{R} + ikct}\right], \tag{80}$$

$$\vec{\boldsymbol{A}}(\vec{r},t) = \left(\frac{\hbar \mathrm{V}}{128\pi^6\varepsilon_{\mathrm{v}}}\right)^{1/2} \iiint_{-\infty}^{\infty} d^3k\, \hat{z}\omega_k{}^{-1/2} \left[\boldsymbol{a}(\vec{k}) e^{i\vec{k}\cdot\vec{R}-ikc} + \boldsymbol{a}^{\dagger}(\vec{k}) e^{-i\vec{k}\cdot\vec{R}+ikct}\right], \tag{81}$$

$$\vec{\boldsymbol{E}}(\vec{r},t) = -i\left(\frac{\hbar \mathrm{V}}{128\pi^6\varepsilon_{\mathrm{v}}}\right)^{1/2} \iiint_{-\infty}^{\infty} d^3k \left[\hat{k}\times(\hat{k}\times\hat{z})\right]\omega_k{}^{1/2} \left[\boldsymbol{a}(\vec{k}) e^{i\vec{k}\cdot\vec{R}-ikct} - \boldsymbol{a}^{\dagger}(\vec{k}) e^{-i\vec{k}\cdot\vec{R}+ikct}\right], \tag{82}$$

$$\vec{\boldsymbol{B}}(\vec{r},t) = i\left(\frac{\hbar \mathrm{V}}{128\pi^6\varepsilon_{\mathrm{v}}}\right)^{1/2} \iiint_{-\infty}^{\infty} d^3k\, k(\hat{k}\times\hat{z})\omega_k{}^{-1/2} \left[\boldsymbol{a}(\vec{k}) e^{i\vec{k}\cdot\vec{R}-ikct} - \boldsymbol{a}^{\dagger}(\vec{k}) e^{-i\vec{k}\cdot\vec{R}+ikct}\right], \tag{83}$$

$$\boldsymbol{H} = \frac{1}{2}\frac{\mathrm{V}}{8\pi^3} \iiint_{-\infty}^{\infty} d^3k\, \hbar\omega_k \sin^2\varkappa_{\vec{k}} \left[\boldsymbol{a}(\vec{k})\boldsymbol{a}^{\dagger}(\vec{k}) + \boldsymbol{a}^{\dagger}(\vec{k})\boldsymbol{a}(\vec{k})\right] = \frac{\mathrm{V}}{8\pi^3} \iiint_{-\infty}^{\infty} d^3k\, \hbar\omega_k \sin^2\varkappa_{\vec{k}} \left[\boldsymbol{a}^{\dagger}(\vec{k})\boldsymbol{a}(\vec{k}) + \frac{1}{2}\right], \tag{84}$$

$$\vec{\boldsymbol{G}} = \frac{1}{2}\frac{\mathrm{V}}{8\pi^3} \iiint_{-\infty}^{\infty} d^3k\, \hbar\vec{k} \sin^2\varkappa_{\vec{k}} \left[\boldsymbol{a}(\vec{k})\boldsymbol{a}^{\dagger}(\vec{k}) + \boldsymbol{a}^{\dagger}(\vec{k})\boldsymbol{a}(\vec{k})\right] = \frac{\mathrm{V}}{8\pi^3} \iiint_{-\infty}^{\infty} d^3k\, \hbar\vec{k} \sin^2\varkappa_{\vec{k}} \left[\boldsymbol{a}^{\dagger}(\vec{k})\boldsymbol{a}(\vec{k}) + \frac{1}{2}\right]. \tag{85}$$

The $\sin^2\varkappa_{\vec{k}}$ appearing in the Hamiltonian, the linear electromagnetic momentum, and the Poynting vector operators give the direction $\vec{k}$ relative to the direction of the emitting dipole for each field mode, such that EM field emission is maximal for direction perpendicular to the emitting dipole and zero along its direction, in agreement with classical electromagnetism.

## 5. Discussion

### 5.1. Consequences of decoupling the potentials and fields from the charges generating them

The QED approach to quantization of the electromagnetic field introduced by Dirac [1] has led to many useful insights that are in excellent agreement with experiments [4,6], and it ought to be recognized as the result of great physical insight and mathematical intuition. It, nevertheless, needs to be carefully evaluated when applied to single dipole emitters, for which the connection between the source and its radiated field needs to be considered, as discussed next in a broader context.

The choice of gauge should leave the EM field unaffected, since one can choose a scalar quantity $\chi$ that allows for the original scalar and vector potentials $\phi$ and $\vec{A}$ (in the Lorenz gauge) to be changed to

$$\phi' = \phi - \frac{\partial\chi}{\partial t}, \tag{86}$$

$$\vec{A}' = \vec{A} + \vec{\nabla}\chi. \tag{87}$$

For the dipole radiator introduced in section 2, the original $\phi$ and $\vec{A}$, which obey the Lorenz condition [21], produce electric and magnetic fields given by Eqns. (63) and (64), which are in agreement with the classical dipole radiation pattern, i.e., the fields are zero for wave vectors oriented along the dipole and reach their maxima for orientations perpendicular to the dipole. The standard QED approach is essentially to choose $\chi = \int_{-\infty}^{t} dT\,\phi(\vec{r},T)$ in Eqns. (86) and (87) such that the scalar potential becomes zero while the vector potential becomes

$$\vec{A}' = \vec{A} + \int_{-\infty}^{t} d\tau \vec{\nabla}\phi(\vec{r},\tau). \tag{88}$$

To be precise, this is in fact the Hamiltonian or temporal gauge, which is only equivalent to the Coulomb gauge under certain conditions [28]. The fact that the expressions of the EM fields given by equations (30) and (31), which are basically the same as those used in the standard QED treatment, do not follow the classical radiation pattern is due to setting the scalar potential to zero without adding $\vec{\nabla}\chi$ to the vector potential to preserve the EM field expressions derived under the Lorenz gauge. In other words, our approach in section 3 and, hence, the standard QED, relied on the condition

$$\nabla \cdot \vec{A}' = 0, \tag{89}$$

instead of

$$\nabla \cdot \vec{A}' = \frac{1}{\varepsilon_{\mathrm{v}}} \int_{-\infty}^{t} d\tau \rho(\vec{r},\tau), \tag{90}$$

and assumed that the expansion coefficients, $\mathcal{A}(\vec{k})$, in the k-form of the vector potential, given by Eqn. (29), are constant [5]. Those assumptions led to the orthogonality relation, Eqn. (40), which was used to cancel out extra terms contributed by the magnetic field to the electromagnetic Hamiltonian, Eqn. (39) but could not compensate for the disappearance of velocity-dependent terms multiplied by $\left[1 - \left(\hat{k} \cdot \hat{z}\right)^2\right]$ from the Hamiltonian [compare Eqn. (39) to (70)]. Thus, even though the dipole may cease to exist at the time, *t*, of observation of its emitted EM field, its contribution to the field is cumulative (i.e., it is integrated over time prior to and up to *t*) and cannot be simply set to zero; that is, one needs to use condition (90), instead of (89), which does not lead to the same orthogonality relation as (40).

Using the present framework instead of the standard implementation of the Coulomb gauge, we showed in section 4 that the quantum Hamiltonian and momentum operators acquire explicit dependences on the polar angle $\varkappa_{\vec{k}}$ made by the direction of each field mode, $\vec{k}$, with the orientation of the emitting dipole, $\hat{z}$, as expressed by Eqns. (84) and (85). In the discrete case [i.e., when using the transformation (53)], the Hamiltonian and linear momentum operators,

$$\boldsymbol{H} = \frac{1}{2}\sum_{\vec{k}} \hbar\omega_k \sin^2 \varkappa_{\vec{k}} \left[\boldsymbol{a}(\vec{k})\boldsymbol{a}^\dagger(\vec{k}) + \boldsymbol{a}^\dagger(\vec{k})\boldsymbol{a}(\vec{k})\right] = \sum_{\vec{k}} \hbar\omega_k \sin^2 \varkappa_{\vec{k}} \left[\boldsymbol{a}^\dagger(\vec{k})\boldsymbol{a}(\vec{k}) + \frac{1}{2}\right], \tag{91}$$

$$\vec{\boldsymbol{G}} = \frac{1}{2}\sum_{\vec{k}} \hbar\vec{k} \sin^2 \varkappa_{\vec{k}} \left[\boldsymbol{a}(\vec{k})\boldsymbol{a}^\dagger(\vec{k}) + \boldsymbol{a}^\dagger(\vec{k})\boldsymbol{a}(\vec{k})\right] = \sum_{\vec{k}} \hbar\vec{k} \sin^2 \varkappa_{\vec{k}} \left[\boldsymbol{a}^\dagger(\vec{k})\boldsymbol{a}(\vec{k}) + \frac{1}{2}\right]. \tag{92}$$

give the energy and momentum eigenvalues

$$H = \sum_{\vec{k}} \hbar\omega_k \sin^2 \varkappa_{\vec{k}} \left[1 + \frac{1}{2}\right], \tag{93}$$

$$\vec{G} = \sum_{\vec{k}} \hbar\vec{k} \sin^2 \varkappa_{\vec{k}} \left[1 + \frac{1}{2}\right], \tag{94}$$

where the $\sin^2 \varkappa_{\vec{k}}$ factor ensures that there is no radiation emission along the direction of the dipole moment and thereby restores the agreement with the classical dipole radiation pattern [23,24]. We took it as the probability of photon emission rather than a modifier of the energy of each mode, in keeping with the spirit of quantum mechanics, and since there is currently no clear evidence that excited atoms or molecules emit photons with different wavelengths for different transition dipole orientations, except when orientation affects the accessibility of different vibronic modes within highly restrictive environments [29,30].

### 5.2. Comparison of the creation and annihilation operator expressions to those obtained from Heisenberg's equation of motion

In Dirac's original approach to quantization of the EM field, the plane wave amplitudes in the plane wave expansion formula, which are subsequently replaced by creation and annihilation operators, are not related to any properties of the oscillator, since the connection between the emitter and the emitted field is cut "by design," i.e., in order to simplify the derivations and obtain expressions that are in agreement with experiment (see above). However, expressions for each of the two operators may be obtained by solving the Heisenberg equations of motion for a dipole oscillator reduced to a single oscillating electron, as discussed in Chapter 2 of Prof. Peter Milonni's highly pedagogical book on quantum vacuum [14]. Those results define each of the two operators as the sum of two terms, which, employing Eqns. (20), (21), (49), and (50) (with the nucleus contributions removed and the strong small-dipole approximation made for conformity), may be written as

$$\boldsymbol{a}(\vec{k}) = \boldsymbol{a}_{\mathrm{v}} - ie\frac{1}{(2\mathrm{V}\varepsilon_{\mathrm{v}}\hbar\omega_k)^{1/2}}\int_{-\infty}^{t} d\tau \Theta_0(\tau)\overrightarrow{\mathrm{v}_-}(\tau)e^{ikc\tau} \tag{95}$$

and

$$\boldsymbol{a}^{\dagger}(\vec{k}) = \boldsymbol{a}^{\dagger}{}_{\mathrm{v}} + ie\frac{1}{(2\mathrm{V}\varepsilon_{\mathrm{v}}\hbar\omega_k)^{1/2}}\int_{-\infty}^{t} d\tau\Theta_0(\tau)\overrightarrow{\mathrm{v}_{-}}(\tau)e^{-ik} \quad , \tag{96}$$

respectively. The first terms, which are added by hand in this paper, have been previously associated with the zero-point, or vacuum, fluctuations (i.e., quantum noise), which enter the Heisenberg equations of motion [but not our classical electrodynamics expressions given by Eqns. (20) and (21)] via commutation relations [14]. The constant coefficients in front of the integral are identical to those in the above reference if provision is the latter are converted to SI units. Note also that there is no dependence on the angle between the dipole and the k vector in these expressions, confirming the premise of this report.

Also interesting, by not assuming that the physical quantities and constants in the classical Eqns. (20) and (21) take the same values in the quantization conditions expressed by Eqns. (49) and (50), the constant coefficients in front of the integrals in (95) and (96) are rederived, after indexing the latter with subscript "q," as

$$\frac{e}{16\pi^3\varepsilon_{\mathrm{v}}kc}\left(\frac{128\pi^6\varepsilon_{\mathrm{v,q}}k_q c_q}{\hbar\mathrm{V}_q}\right)^{1/2} = \frac{e}{e_q}\frac{\varepsilon_{\mathrm{v,q}}}{\varepsilon_{\mathrm{v}}}\frac{c_q}{c}\left(\frac{\lambda^2}{\lambda_q\mathrm{V}_q}\right)^{1/2}\alpha^{1/2},$$

where we also used $k = 2\pi/\lambda$, $k_q = 2\pi/\lambda_q$, and $\hbar = e_q{}^2/(4\pi\varepsilon_{\mathrm{v,q}}c_q\alpha)$, with $\alpha$ being the fine structure constant. It is now clear that the mathematical procedure of quantization involves making one of the ratios or (a combination of them) in the last expression less than unity by $\alpha^{1/2} \approx 11.7$. Regardless of specifically which physical constant takes a different value at quantum level, the net result is an increased strength of electromagnetic interactions, in agreement with Feynman's QED.

### 5.3. Stimulated and spontaneous emission

It is generally understood that each term indexed by $k$ in equations (94) and (95) does not represent a photon, but rather a single mode associated with radiation emission [8]. Since we assumed a single radiator (i.e., oscillating dipole) which emitted its electromagnetic energy by the time of measurement, the sum over all modes is equivalent to the energy of the entire photon, plus the vacuum energy. To understand how this relates to the well-known expression $E = \hbar\omega$, we need to bring up the process of measurement.

We will first need an appropriate light source. Let us consider a continuous wave (CW) laser cavity wherein the modes that are not colinear with the optical axis of the cavity are optically excluded, because they are not reflected by the cavity mirrors and, therefore, not amplified. Of those modes that do get reflected, only the ones with frequencies matching the optical transition of the laser gain medium ($\overrightarrow{k_s}$) are amplified and become the output of the laser. Using a filter or some other method to attenuate the laser's output, the photons that enter our measurement system, where the dipole radiator of interest resides, are separated by time intervals longer than the time needed for all the physical processes of interest here to complete. Under these conditions, our light source itself emits EM fields given by equations (82) and (83) in a direction perpendicular to the direction of the selected laser dipole (i.e., $\hat{k}_s \perp \hat{z}$ and, therefore, $\varkappa_{\overrightarrow{k_s}} = \frac{\pi}{2}$), in which the wave vector distribution $(\omega_k/c)$ in the integrals is given by the Dirac delta function $\frac{8\pi^3}{\mathrm{V}} k\delta\left(\vec{k} - \overrightarrow{k_s}\right)$. Thus, we obtain the fields

$$\vec{\boldsymbol{E}}_s(\vec{r},t) = i\left(\frac{\hbar}{2\varepsilon_{\mathrm{v}}\mathrm{V}}\right)^{1/2} \hat{z}{\omega_s}^{1/2}\left[\boldsymbol{a}\left(\vec{k}_s\right)e^{i\vec{k}_s\cdot\vec{R}-ik_sct} - \boldsymbol{a}^{\dagger}\left(\vec{k}_s\right)e^{-i\vec{k}_s\cdot\vec{R}+ik_sct}\right], \tag{97}$$

$$\vec{\boldsymbol{B}}_s(\vec{r},t) = i\left(\frac{\hbar}{2\varepsilon_{\mathrm{v}}\mathrm{V}}\right)^{1/2} k_s\left(\hat{k}_s \times \hat{z}\right){\omega_s}^{-1/2}\left[\boldsymbol{a}\left(\vec{k}_s\right)e^{i\vec{k}_s\cdot\vec{R}-ik_sct} - \boldsymbol{a}^{\dagger}\left(\vec{k}_s\right)e^{-i\vec{k}_s\cdot\vec{R}+ik_sct}\right], \tag{98}$$

which correspond to a single mode characterized by the Hamiltonian

$$\boldsymbol{H}_s = \hbar\omega_s\left[\boldsymbol{a}^{\dagger}\left(\vec{k}_s\right)\boldsymbol{a}\left(\vec{k}_s\right) + \frac{1}{2}\right] \tag{100}$$

as obtained from Eqn. (84), and the electromagnetic momentum operator

$$\vec{\boldsymbol{G}}_s = \hbar\vec{k}\left[\boldsymbol{a}^{\dagger}\left(\vec{k}_s\right)\boldsymbol{a}\left(\vec{k}_s\right) + \frac{1}{2}\right], \tag{101}$$

obtained from Eqn. (85).

We can use this source to investigate the process of stimulated emission by an excited dipole radiator consisting of the same type of atom as the ones in the gain medium of the laser. For simplicity, we ignore the fact that the stimulating field may modify the behavior of the dipole, which in its turn could make additional contributions to the emitted field, without qualitatively changing the discussion herein. The combined stimulating (subscript "s") fields and the fields emitted by the already excited dipole (subscript "d"), $\vec{E}_t = \vec{E}_s + \vec{E}_d$, and $\vec{B}_t = \vec{B}_s + \vec{B}_d$, carry electromagnetic energy integrated over the entire space centered around the dipole, of which we are only interested in the interference term between the stimulating fields, given by Eqns. (82) and (83), and the dipole fields, defined by Eqns. (98) and (99). Using the procedures employed above, the Hamiltonian and momentum operators corresponding to the interference terms are derived as (see Appendix G)

$$\boldsymbol{H}_{SE} = \varepsilon_{\mathrm{v}}\iiint_{-\infty}^{\infty} d^3R\left(\vec{\boldsymbol{E}}_s \cdot \vec{\boldsymbol{E}}_d\right) + \frac{1}{\mu_{\mathrm{v}}}\iiint_{-\infty}^{\infty} d^3R\left(\vec{\boldsymbol{B}}_s \cdot \vec{\boldsymbol{B}}_d\right) = 2\hbar\omega_s\sin^2\varkappa_{\vec{k}S}\left[\boldsymbol{a}^{\dagger}\left(\vec{k}_s\right)\boldsymbol{a}\left(\vec{k}_s\right) + \frac{1}{2}\right], \tag{102}$$

and, respectively,

$$\vec{\boldsymbol{G}}_{SE} = \varepsilon_{\mathrm{v}} \iiint_{-\infty}^{\infty} d^3R \left(\vec{\boldsymbol{E}}_s \times \vec{\boldsymbol{B}}_d\right) + \varepsilon_{\mathrm{v}} \iiint_{-\infty}^{\infty} d^3R \left(\vec{\boldsymbol{E}}_d \times \vec{\boldsymbol{B}}_s\right) = 2\hbar\vec{k}_s \sin^2 \varkappa_{\vec{k}S} \left[\boldsymbol{a}^{\dagger}\left(\vec{k}_s\right)\boldsymbol{a}\left(\vec{k}_s\right) + \frac{1}{2}\right]. \quad (103)$$

These equations suggest that the stimulating field collapses the electromagnetic radiation emitted by the dipole into a single mode. Together, the dipole and stimulating fields carry away, in the original direction of propagation of the stimulating field, the energy and momentum [represented by the eigenvalues of (102) and (103)] of the electromagnetic radiation which, compared to Eqns. (100) and (101), are equivalent to those for two photons.

Since it is known that spontaneous emission causes the emitting atom or molecule to recoil and that vacuum fluctuations are at least partly responsible for spontaneous emission [14], it is very tempting to assume that vacuum fluctuations, possessing just the right phase, can interfere constructively with the dipole radiation to collapse it into a single mode (or a few modes with a relatively narrow frequency distribution), just as we concluded that stimulated emission does. If it does happen, this collapse too must obey the $\sin^2 \varkappa_{\vec{k}}$ rule predicted by equations (98) and (99), which could be tested experimentally.

## 6. Conclusions

Introduction of plane wave expansions of electromagnetic fields starting from distributions of charges and currents as presented in this paper indicates that the spatial modes of single dipolar emitters follow an angular probability distribution that agrees with the classical dipole radiation pattern, and it predicts that individual photons are emitted into single (or just a few) spatial modes under stimulation by light or vacuum fluctuations.

The present interpretation of $\sin^2 \varkappa_{\vec{k}}$ as a probability of photon emission, while being in agreement with classical electrodynamics and supported by single-dipole-imaging experiments, remains to be validated by direct experimental testing. An experimental setup will be proposed elsewhere, which could allow for determination of ***(i)*** the $\sin^2 \varkappa_{\vec{k}}$ dependence of the photon emission probability (i.e., relative intensity) and ***(ii)*** the emission spectrum for each orientation of the transition dipole of fluorescent molecules whose dipole orientation is optically selected by choosing the linear polarization orientation of excitation light.

Further development of this theoretical work could more deeply explore the connection between the conjugate momentum and position variables with the dipole properties, via Eqns. (20) and (21), as well as  the quantum mechanics of the atom. A protocol may also be established for incorporating vacuum fluctuations, which are inherent in the commutation relations, into the classical electrodynamics results introduced in section 3 by modifying the boundary conditions used for deriving expressions to replace Eqns. (1) and (2) [21]. Finally, it should be possible in principle to study radiation emission from excited atoms in the presence and absence of vacuum fluctuations or other stimulating fields by replacing the dipole model with one that incorporates numerical superposition of excited-state and ground-state orbitals [7], to visualize any differences in radiation emission patterns in the presence and absence of stimulation.

On an immediate practical level, the present study allows for interpretation of the results of single-molecular dipole imaging, the use of spontaneous and stimulated emission in quantitative optical imaging [13,19,31,32], and the interpretation of a plethora of interesting, yet puzzling, single-photon level experiments (see, e.g., chapter 6 in [8] and references therein). In addition, in conjunction with FRET spectrometry [13], this study may lead to fully quantitative determination of the structure of supramolecular complexes without making assumptions regarding the orientation of the transitions dipoles within a complex (such as cylindrical averaging [25]) or performing costly and lengthy computer simulations to incorporate such orientations within the theoretical models [12].

## Acknowledgements

I am indebted to Prof. Peter Milonni for very helpful conversations and critical reading of the manuscript, and to Fr. John Konkle for support and encouragement. This work has been partly supported through a Discovery and Innovation Grant from the University of Wisconsin-Milwaukee (DIG 101X453).

## Conflict of Interest

The author declares that there are no competing financial interests.

## Data Availability Statement

Data sharing is not applicable to this article, as no new data were created or analyzed in this study.

## Quantization of the electromagnetic fields from single atomic or molecular radiators

Valerică Raicu

Department of Physics and Astronomy, University of Wisconsin-Milwaukee, Milwaukee, WI 53211, USA

Email: vraicu@uwm.edu

## Supporting Information

### Appendix A: Derivation of equations (3) and (4) for the scalar and vector potential

The integral $\iiint_{-\infty}^{\infty} \frac{1}{k} \sin[ck(t-\tau)] \sin(\vec{k}\cdot\vec{R})\, d^3k$ may be written in spherical coordinates in $\vec{k}$, by replacing $\vec{k}\cdot(\vec{r}-\vec{r'}) \equiv \vec{k}\cdot\vec{R}$ with $kR\cos\theta$, as

$$\iiint_{-\infty}^{\infty} \frac{1}{k} \sin[ck(t-\tau)] \sin(\vec{k}\cdot\vec{R})\, d^3k = \int_0^{2\pi} d\varphi \int_0^{\pi} \sin\theta\, d\theta \int_0^{\infty} \sin[ck(t-\tau)] \sin(kR\cos\theta)\, kdk, \tag{A1}$$

after using the substitutions $u = \cos\theta$ and $du = -\sin\theta\, d\theta$ and integrating, (A1) becomes:

$$\iiint_{-\infty}^{\infty} d^3k \frac{1}{k} \sin[ck(t-\tau)] \sin[kR(\tau)\cos\theta] = 0. \tag{A2}$$

Thus, we may subtract this integral from the integrals over $k$ in Eqns. (1) and (2) without changing the final result, and obtain

$$\phi(\vec{r},t) = \frac{c}{8\pi^3\varepsilon_{\text{v}}}\int_{-\infty}^{t} d\tau \iiint_{-\infty}^{\infty} d^3r'\rho\left(\vec{r'},\tau\right)\Theta_0(\tau)\iiint_{-\infty}^{\infty} d^3k\frac{1}{k}\left\{\sin[ck(t-\tau)]\cos\left(\vec{k}\cdot\vec{R}\right) - \cos[ck(t-\tau)]\sin\left(\vec{k}\cdot\vec{R}\right)\right\}, \tag{A3}$$

$$\vec{A}(\vec{r},t) = \frac{c\mu_{\text{v}}}{8\pi^3}\int_{-\infty}^{t} d\tau \iiint_{-\infty}^{\infty} d^3r'\vec{J}\left(\vec{r'},\tau\right)\Theta_0(\tau)\iiint_{-\infty}^{\infty} d^3k\frac{1}{k}\left\{\sin[ck(t-\tau)]\cos\left(\vec{k}\cdot\vec{R}\right) - \cos[ck(t-\tau)]\sin\left(\vec{k}\cdot\vec{R}\right)\right\}. \tag{A4}$$

After using the trigonometric identity $\sin(a-b) = \sin a\cos b - \cos a\sin b$, we finally obtain:

$$\phi(\vec{r},t) = \frac{c}{8\pi^3\varepsilon_{\text{v}}}\int_{-\infty}^{t} d\tau \iiint_{-\infty}^{\infty} d^3r'\rho\left(\vec{r'},\tau\right)\Theta_0(\tau)\iiint_{-\infty}^{\infty} d^3k\frac{1}{k}\sin\left[kc(t-\tau) - \vec{k}\cdot\vec{R}\right], \tag{A5}$$

$$\vec{A}(\vec{r},t) = \frac{c\mu_{\text{v}}}{8\pi^3}\int_{-\infty}^{t} d\tau \iiint_{-\infty}^{\infty} d^3r'\vec{J}\left(\vec{r'},\tau\right)\Theta_0(\tau)\iiint_{-\infty}^{\infty} d^3k\frac{1}{k}\sin\left[kc(t-\tau) - \vec{k}\cdot\vec{R}\right]. \tag{A6}$$

## Appendix B: Derivation of equations (7) and (8) in the main text

Changing the order of integration between $r'$ and $k$ in equations

$$\phi(\vec{r},t) = i\frac{c}{16\pi^3\varepsilon_{\mathrm{v}}}\iiint_{-\infty}^{\infty} d^3k\frac{1}{k}\int_{-\infty}^{t} d\tau\Theta_0(\tau)\iiint_{-\infty}^{\infty} d^3r'\rho\left(\vec{r'},\tau\right)e^{i\vec{k}\cdot\left(\vec{r}-\vec{r'}\right)-ik\ (t-\tau)} -$$
$$i\frac{c}{16\pi^3\varepsilon_{\mathrm{v}}}\iiint_{-\infty}^{\infty} d^3k\frac{1}{k}\int_{-\infty}^{t} d\tau\Theta_0(\tau)\iiint_{-\infty}^{\infty} d^3r'\rho\left(\vec{r'},\tau\right)e^{-i\vec{k}\cdot\left(\vec{r}-\vec{r'}\right)+ikc(t-\tau)}, \qquad (5)$$

and

$$\vec{A}(\vec{r},t) = i\frac{c\mu_{\mathrm{v}}}{16\pi^3}\iiint_{-\infty}^{\infty} d^3k\frac{1}{k}\int_{-\infty}^{t} d\tau\Theta_0(\tau)\iiint_{-\infty}^{\infty} d^3r'\vec{j}\left(\vec{r'},\tau\right)e^{i\vec{k}\cdot\left(\vec{r}-\vec{r'}\right)-ik\ (t-\tau)} -$$
$$i\frac{c\mu_{\mathrm{v}}}{16\pi^3}\iiint_{-\infty}^{\infty} d^3k\frac{1}{k}\int_{-\infty}^{t} d\tau\Theta_0(\tau)\iiint_{-\infty}^{\infty} d^3r'\vec{j}\left(\vec{r'},\tau\right)e^{-i\vec{k}\cdot\left(\vec{r}-\vec{r'}\right)+ikc(t-\tau)}, \qquad (6)$$

from the main text, we have

$$\phi(\vec{r},t) = i\frac{c}{16\pi^3\varepsilon_{\mathrm{v}}}\int_{-\infty}^{t} d\tau\Theta_0(\tau)\iiint_{-\infty}^{\infty} d^3r'\rho\left(\vec{r'},\tau\right)\iiint_{-\infty}^{\infty} d^3k\frac{1}{k}e^{i\vec{k}\cdot\left(\vec{r}-\vec{r'}\right)-ikc(t-\tau)} -$$
$$i\frac{c}{16\pi^3\varepsilon_{\mathrm{v}}}\int_{-\infty}^{t} d\tau\Theta_0(\tau)\iiint_{-\infty}^{\infty} d^3r'\rho\left(\vec{r'},\tau\right)\iiint_{-\infty}^{\infty} d^3k\frac{1}{k}e^{-i\vec{k}\cdot\left(\vec{r}-\vec{r'}\right)+ikc(t-\tau)}, \qquad (B1)$$

$$\vec{A}(\vec{r},t) = i\frac{c\mu_{\mathrm{v}}}{16\pi^3}\int_{-\infty}^{t} d\tau\Theta_0(\tau)\iiint_{-\infty}^{\infty} d^3r'\vec{j}\left(\vec{r'},\tau\right)\iiint_{-\infty}^{\infty} d^3k\frac{1}{k}e^{i\vec{k}\cdot\left(\vec{r}-\vec{r'}\right)-ikc(t-\tau)} -$$
$$i\frac{c\mu_{\mathrm{v}}}{16\pi^3}\int_{-\infty}^{t} d\tau\Theta_0(\tau)\iiint_{-\infty}^{\infty} d^3r'\vec{j}\left(\vec{r'},\tau\right)\iiint_{-\infty}^{\infty} d^3k\frac{1}{k}e^{-i\vec{k}\cdot\left(\vec{r}-\vec{r'}\right)+ikc(t-\tau)}. \qquad (B2)$$

Switching to spherical coordinates in $\vec{k}$ by replacing $\vec{k}\cdot\left(\vec{r}-\vec{r'}\right)$ with $k\left|\vec{r}-\vec{r'}\right|\cos\theta$, the first integrals with respect to $k$ become

$$\iiint_{-\infty}^{\infty} d^3k\frac{1}{k}e^{i\vec{k}\cdot\left(\vec{r}-\vec{r'}\right)-ikc(t-\tau)} = \int_0^{2\pi} d\varphi\int_0^{\infty} dk\int_0^{\pi} d\theta\sin\theta\, e^{ik\left|\vec{r}-\vec{r'}\right|\cos\theta-ikc(t-\tau)}k =$$
$$-\frac{2\pi}{i\left|\vec{r}-\vec{r'}\right|}\int_0^{\infty} dk e^{-ik\left|\vec{r}-\vec{r'}\right|-ikc(t-\tau)} + \frac{2\pi}{i\left|\vec{r}-\vec{r'}\right|}\int_0^{\infty} dk e^{ik\left|\vec{r}-\vec{r'}\right|-ikc(t-\tau)}, \qquad (B3)$$

while the second one becomes

$$\iiint_{-\infty}^{\infty} d^3k\frac{1}{k}e^{-i\vec{k}\cdot\left(\vec{r}-\vec{r'}\right)+ikc(t-\tau)} = \int_0^{2\pi} d\varphi\int_0^{\infty} dk\int_0^{\pi} d\theta\sin\theta\, e^{-ik\left|\vec{r}-\vec{r'}\right|\cos\theta+ikc(t-\tau)}k =$$
$$\frac{2\pi}{i\left|\vec{r}-\vec{r'}\right|}\int_0^{\infty} dk e^{ik\left|\vec{r}-\vec{r'}\right|+ikc(t-\tau)} - \frac{2\pi}{i\left|\vec{r}-\vec{r'}\right|}\int_0^{\infty} dk e^{-ik\left|\vec{r}-\vec{r'}\right|+ikc(t-\tau)}, \qquad (B4)$$

where we used a simple color code to track the origin of the delta functions in the finale expressions shown at the end of this appendix.

Using Eqns. (B3) and (B4) to replace the integrals with respect to $k$ in (B1) and (B2), we obtain

$$\phi(\vec{r},t) = -\frac{c}{8\pi^2\varepsilon_{\mathrm{v}}}\int_{-\infty}^{t} d\tau\Theta_0(\tau)\iiint_{-\infty}^{\infty} d^3r'\frac{\rho\left(\vec{r'},\tau\right)}{\left|\vec{r}-\vec{r'}\right|}\int_0^{\infty} dk e^{-ik\left|\vec{r}-\vec{r'}\right|-ikc(t-\tau)} +$$
$$\frac{c}{8\pi^2\varepsilon_{\mathrm{v}}}\int_{-\infty}^{t} d\tau\Theta_0(\tau)\iiint_{-\infty}^{\infty} d^3r'\frac{\rho\left(\vec{r'},\tau\right)}{\left|\vec{r}-\vec{r'}\right|}\int_0^{\infty} dk e^{ik\left|\vec{r}-\vec{r'}\right|-ikc(t-\tau)} -$$

$$\frac{c}{8\pi^2\varepsilon_{\mathrm{v}}}\int_{-\infty}^{t} d\tau\Theta_0(\tau)\iiint_{-\infty}^{\infty} d^3r' \frac{\rho(\vec{r'},\tau)}{|\vec{r}-\vec{r'}|}\int_0^\infty dk e^{ik|\vec{r}-\vec{r'}|+ikc(t-\tau)} +$$

$$\frac{c}{8\pi^2\varepsilon_{\mathrm{v}}}\int_{-\infty}^{t} d\tau\Theta_0(\tau)\iiint_{-\infty}^{\infty} d^3r' \frac{\rho(\vec{r'},\tau)}{|\vec{r}-\vec{r'}|}\int_0^\infty dk e^{-i\ |\vec{r}-\vec{r'}|+ik\ (t-\tau)}, \tag{B5}$$

$$\vec{A}(\vec{r},t) = -\frac{c\mu_{\mathrm{v}}}{8\pi^2}\int_{-\infty}^{t} d\tau\Theta_0(\tau)\iiint_{-\infty}^{\infty} d^3r' \frac{\vec{j}(\vec{r'},\tau)}{|\vec{r}-\vec{r'}|}\int_0^\infty dk e^{-ik|\vec{r}-\vec{r'}|-ikc(t-\tau)} +$$

$$\frac{c\mu_{\mathrm{v}}}{8\pi^2}\int_{-\infty}^{t} d\tau\Theta_0(\tau)\iiint_{-\infty}^{\infty} d^3r' \frac{\vec{j}(\vec{r'},\tau)}{|\vec{r}-\vec{r'}|}\int_0^\infty dk e^{-i\ |\vec{r}-\vec{r'}|-ikc(t-\tau)} -$$

$$\frac{c\mu_{\mathrm{v}}}{8\pi^2}\int_{-\infty}^{t} d\tau\Theta_0(\tau)\iiint_{-\infty}^{\infty} d^3r' \frac{\vec{j}(\vec{r'},\tau)}{|\vec{r}-\vec{r'}|}\int_0^\infty dk e^{ik|\vec{r}-\vec{r'}|+ikc(t-\tau)} +$$

$$\frac{c\mu_{\mathrm{v}}}{8\pi^2}\int_{-\infty}^{t} d\tau\Theta_0(\tau)\iiint_{-\infty}^{\infty} d^3r' \frac{\vec{j}(\vec{r'},\tau)}{|\vec{r}-\vec{r'}|}\int_0^\infty dk e^{-ik|\vec{r}-\vec{r'}|+ikc(t-\tau)}, \tag{B6}$$

or, after grouping the second with the fourth exponentials and the first with the third one and using Euler's formula,

$$\phi(\vec{r},t) = \frac{c}{8\pi^2\varepsilon_{\mathrm{v}}}\int_{-\infty}^{t} d\tau\Theta_0(\tau)\iiint_{-\infty}^{\infty} d^3r' \frac{\rho(\vec{r'},\tau)}{|\vec{r}-\vec{r'}|}\int_0^\infty dk \cos\left(kc\tau - kct + k|\vec{r}-\vec{r'}|\right) -$$

$$\frac{c}{4\pi^2\varepsilon_{\mathrm{v}}}\int_{-\infty}^{t} d\tau\Theta_0(\tau)\iiint_{-\infty}^{\infty} d^3r' \frac{\rho(\vec{r'},\tau)}{|\vec{r}-\vec{r'}|}\int_0^\infty dk \cos\left(kc\tau - kct - k|\vec{r}-\vec{r'}|\right), \tag{B8}$$

$$\vec{A}(\vec{r},t) = \frac{c\mu_{\mathrm{v}}}{8\pi^2}\int_{-\infty}^{t} d\tau\Theta_0(\tau)\iiint_{-\infty}^{\infty} d^3r' \frac{\vec{j}(\vec{r'},\tau)}{|\vec{r}-\vec{r'}|}\int_0^\infty dk \cos\left(kc\tau - kct + k|\vec{r}-\vec{r'}|\right) -$$

$$\frac{c\mu_{\mathrm{v}}}{8\pi^2}\int_{-\infty}^{t} d\tau\Theta_0(\tau)\iiint_{-\infty}^{\infty} d^3r' \frac{\vec{j}(\vec{r'},\tau)}{|\vec{r}-\vec{r'}|}\int_0^\infty dk \cos\left(kc\tau - kct - k|\vec{r}-\vec{r'}|\right). \tag{B9}$$

But the integrals containing the cosines are related to the Dirac delta functions via $\int_0^\infty dk \cos(kx) = \pi\delta(x)$. With this, and using known properties of the delta function, the last two equations become

$$\phi(\vec{r},t) = \frac{1}{4\pi\varepsilon_{\mathrm{v}}}\iiint_{-\infty}^{\infty} d^3r' \int_{-\infty}^{t} d\tau\Theta_0(\tau)\frac{\rho(\vec{r'},\tau)}{|\vec{r}-\vec{r'}|}\delta\left[\tau - \left(t - \frac{|\vec{r}-\vec{r'}|}{c}\right)\right] -$$

$$\frac{1}{4\pi\varepsilon_{\mathrm{v}}}\iiint_{-\infty}^{\infty} d^3r' \int_{-\infty}^{t} d\tau\Theta_0(\tau)\frac{\rho(\vec{r'},\tau)}{|\vec{r}-\vec{r'}|}\delta\left[\tau - \left(t + \frac{|\vec{r}-\vec{r'}|}{c}\right)\right], \tag{B10}$$

$$\vec{A}(\vec{r},t) = \frac{\mu_{\mathrm{v}}}{4\pi}\int_{-\infty}^{t} d\tau\Theta_0(\tau)\iiint_{-\infty}^{\infty} d^3r' \frac{\vec{j}(\vec{r'},\tau)}{|\vec{r}-\vec{r'}|}\delta\left[\tau - \left(t - \frac{|\vec{r}-\vec{r'}|}{c}\right)\right] - \frac{\mu_{\mathrm{v}}}{4\pi}\iiint_{-\infty}^{\infty} d^3r' \int_{-\infty}^{t} d\tau\Theta_0(\tau)\frac{\vec{j}(\vec{r'},\tau)}{|\vec{r}-\vec{r'}|}\delta\left[\tau - \left(t + \frac{|\vec{r}-\vec{r'}|}{c}\right)\right], \tag{B11}$$

where we also changed the order of integration between $r'$ and $\tau$.

## Appendix C: Separation of the k-form of $\phi(\vec{r},t)$ into velocity- and position-dependent terms

Taking the dot product of $\vec{k}$ with $\mathcal{A}(\vec{k},t)$ in the expression

$$\mathcal{A}(\vec{k},t) = i\frac{ec\mu_{\mathrm{v}}}{16\pi^3}\frac{1}{k}\int_{-\infty}^{t} d\tau e^{ikc\tau - i\vec{k}\cdot\overrightarrow{d'_d}(\tau)}\Theta_0(\tau)\left[\overrightarrow{\mathrm{v}_+}(\tau)e^{-i\vec{k}\cdot\overrightarrow{z'_+}(\tau)} - \overrightarrow{\mathrm{v}_-}(\tau)e^{-i\vec{k}\cdot\overrightarrow{z'_-}(\tau)}\right], \tag{20}$$

we get

$$\vec{k}\cdot\mathcal{A}(\vec{k},t) = i\frac{ec\mu_{\mathrm{v}}}{16\pi^3}\frac{1}{k}\int_{-\infty}^{t} d\tau\Theta_0(\tau)e^{ikc\tau - i\vec{k}\cdot\overrightarrow{d'_d}(\tau)}\vec{k}\cdot\left[\overrightarrow{\mathrm{v}_+}(\tau)e^{-i\vec{k}\cdot\overrightarrow{z'_+}(\tau)} - \overrightarrow{\mathrm{v}_-}(\tau)e^{-i\vec{k}\cdot\overrightarrow{z'_-}(\tau)}\right] =$$
$$-\frac{ec\mu_{\mathrm{v}}}{16\pi^3}\frac{1}{k}\int_{0}^{t} d\tau\Theta_0(\tau)e^{ikc\tau - i\vec{k}\cdot\overrightarrow{d'_d}(\tau)}\frac{d}{d\tau}\left[e^{-i\vec{k}\cdot\overrightarrow{z'_+}(\tau)} - e^{-i\vec{k}\cdot\overrightarrow{z'_-}(\tau)}\right]. \tag{C1}$$

Using the notations: $u = \Theta_0(\tau)e^{ikc\tau - i\vec{k}\cdot\overrightarrow{d'_d}(\tau)} \Rightarrow \frac{du}{d\tau} = \delta(\tau - t_0)e^{ikc\tau - \vec{\ }\cdot\overrightarrow{d'_d}(\tau)} + i\Theta_0(\tau)\left[kc - \vec{k}\cdot\overrightarrow{\mathrm{v}_d}(\tau)\right]e^{ikc\tau - i\vec{k}\cdot\overrightarrow{d'_d}(\tau)}$, $\frac{dw}{d\tau} = \frac{d}{d\tau}\left[e^{-i\vec{k}\cdot\overrightarrow{z'_+}(\tau)} - e^{-i\vec{k}\cdot\overrightarrow{z'_-}(\tau)}\right] \Rightarrow w = e^{-i\vec{k}\cdot\overrightarrow{z'_+}(\tau)} - e^{-i\vec{k}\cdot\overrightarrow{z'_-}(\tau)}$, integrate (C1) by parts and obtain

$$\vec{k}\cdot\vec{\mathcal{A}}(\vec{k},t) = -\frac{ec\ \ _{\mathrm{v}}}{16\pi^3}\frac{1}{k}e^{ikct - i\vec{k}\cdot\overrightarrow{d'_d}(t)}\left[e^{-i\vec{k}\cdot\overrightarrow{z'_+}(t)} - e^{-i\vec{k}\cdot\overrightarrow{z'_-}(t)}\right] + \frac{ec\mu_{\mathrm{v}}}{16\pi^3}\frac{1}{k}\int_{-\infty}^{t} d\tau\delta(\tau -$$
$$t_0)e^{ikc\tau - \vec{\ }\cdot\overrightarrow{d'_d}(\tau)}\left[e^{-i\vec{k}\cdot\overrightarrow{z'_+}(\tau)} - e^{-i\vec{k}\cdot\overrightarrow{z'_-}(\tau)}\right] + i\frac{ec\mu_{\mathrm{v}}}{16\pi^3}\frac{1}{k}\int_{-\infty}^{t} d\tau\Theta_0(\tau)\left[kc - \vec{k}\cdot\right.$$
$$\left.\overrightarrow{\mathrm{v}_d}(\tau)\right]e^{ikc\tau - \vec{\ }\cdot\overrightarrow{d'_d}(\tau)}\left[e^{-i\vec{k}\cdot\overrightarrow{z'_+}(\tau)} - e^{-i\vec{k}\cdot\overrightarrow{z'_-}(\tau)}\right],$$

or, after using the sifting property of the delta function in the second integral [together with $\overrightarrow{z'_\pm}(t_0) = 0$],

$$\vec{k}\cdot\mathcal{A}(\vec{k},t) = -\frac{ec\mu_{\mathrm{v}}}{16\pi^3}\frac{1}{k}e^{ikct - i\vec{k}\cdot\overrightarrow{d'_d}(t)}\left[e^{-i\vec{k}\cdot\overrightarrow{z'_+}(t)} - e^{-i\vec{k}\cdot\overrightarrow{z'_-}(t)}\right] +$$
$$i\frac{e}{16\pi^3\varepsilon_{\mathrm{v}}}\int_{-\infty}^{t} d\tau\Theta_0(\tau)e^{ikc\tau - i\vec{k}\cdot\overrightarrow{d'_d}(\tau)}\left[e^{-i\vec{k}\cdot\overrightarrow{z'_+}(\tau)} - e^{-i\vec{k}\cdot\overrightarrow{z'_-}(\tau)}\right] - i\frac{ec\mu_{\mathrm{v}}}{16\pi^3}\int_{-\infty}^{t} d\tau\Theta_0(\tau)\left[\hat{k}\cdot\right.$$
$$\left.\overrightarrow{\mathrm{v}_d}(\tau)\right]e^{ikc\tau - i\vec{k}\cdot\overrightarrow{d'_d}(\tau)}\left[e^{-i\vec{k}\cdot\overrightarrow{z'_+}(\tau)} - e^{-i\vec{k}\cdot\overrightarrow{z'_-}(\tau)}\right]. \tag{C2}$$

Substitution of $\mathcal{F}(\vec{k},t)$ given by Eqn. (17) in the main text,

$$\mathcal{F}(\vec{k},t) = i\frac{ec}{16\pi^3\varepsilon_{\mathrm{v}}}\frac{1}{k}\int_{-\infty}^{t} d\tau\Theta_0(\tau)e^{ikc\tau - i\vec{k}\cdot\overrightarrow{d'_d}(t)}\left[e^{-i\vec{k}\cdot\overrightarrow{z'_+}(\tau)} - e^{-i\vec{k}\cdot\overrightarrow{z'_-}(\tau)}\right], \tag{17}$$

into (C2) gives

$$\vec{k}\cdot\mathcal{A}(\vec{k},t) = -\frac{ec\mu_{\mathrm{v}}}{16\pi^3}\frac{1}{k}e^{ikct - i\vec{k}\cdot\overrightarrow{d'_d}(t)}\left[e^{-i\vec{k}\cdot\overrightarrow{z'_+}(t)} - e^{-i\vec{k}\cdot\overrightarrow{z'_-}(t)}\right] - i\frac{ec\mu_{\mathrm{v}}}{16\ \ ^3}\int_{-\infty}^{t} d\tau\Theta_0(\tau)\left[\hat{k}\cdot\right.$$
$$\left.\overrightarrow{\mathrm{v}_d}(\tau)\right]e^{ikc\tau - i\vec{k}\cdot\overrightarrow{z'_d}(\tau)}\left[e^{-i\vec{k}\cdot\overrightarrow{d'_+}(\tau)} - e^{-i\vec{k}\cdot\overrightarrow{z'_-}(\tau)}\right] + \frac{k}{c}\mathcal{F}(\vec{k},t), \tag{C3}$$

or, equivalently,

$$\mathcal{F}(\vec{k},t) = c\hat{k}\cdot\mathcal{A}(\vec{k},t) + \frac{e}{16\ \ ^3\varepsilon_{\mathrm{v}}}\frac{1}{k^2}e^{ikct - i\vec{k}\cdot\overrightarrow{d'_d}(t)}\left[e^{-i\vec{k}\cdot\overrightarrow{z'_+}(t)} - e^{-i\vec{k}\cdot\overrightarrow{z'_-}(t)}\right] + c\hat{k}\cdot\vec{\mathcal{V}}(\vec{k},t), \tag{C4}$$

with

$$\mathcal{V}(\vec{k},t) = i\frac{ec\mu_{\mathrm{v}}}{16\pi^3}\frac{1}{k}\int_{-\infty}^{t} d\tau\Theta_0(\tau)e^{ikc\tau - i\vec{k}\cdot\overrightarrow{d'_d}(\tau)}\overrightarrow{\mathrm{v}_d}(\tau)\left[e^{-i\vec{k}\cdot\overrightarrow{z'_+}(\tau)} - e^{-i\vec{k}\cdot\overrightarrow{z'_-}(\tau)}\right]. \tag{C5}$$

Similarly, taking the dot product of $\vec{k}$ with $\overrightarrow{\mathcal{A}^*}(\vec{k},t)$ in

$$\overrightarrow{\mathcal{A}^*}(\vec{k},t) = -i\frac{ec\mu_{\mathrm{v}}}{16\pi^3}\frac{1}{k}\int_{-\infty}^{t} d\tau\Theta_0(\tau)e^{-ikc\tau+i\vec{k}\cdot\overrightarrow{d_d'}(\tau)}\left[\overrightarrow{\mathrm{v}_+}(\tau)e^{i\vec{k}\cdot\overrightarrow{z_+'}(\tau)} - \overrightarrow{\mathrm{v}_-}(\tau)e^{i\vec{k}\cdot\overrightarrow{z_-'}(\tau)}\right], \tag{21}$$

we get

$$\vec{k}\cdot\overrightarrow{\mathcal{A}^*}(\vec{k},t) = -i\frac{ec\mu_{\mathrm{v}}}{16\pi^3}\frac{1}{k}\int_{-\infty}^{t} d\tau\Theta_0(\tau)e^{-ikc\tau+i\vec{k}\cdot\overrightarrow{d_d'}(\tau)}\vec{k}\cdot\left[\overrightarrow{\mathrm{v}_+}(\tau)e^{i\vec{k}\cdot\overrightarrow{z_+'}(\tau)} - \overrightarrow{\mathrm{v}_-}(\tau)e^{i\vec{k}\cdot\overrightarrow{z_-'}(\tau)}\right] = -\frac{ec\mu_{\mathrm{v}}}{16\pi^3}\frac{1}{k}\int_{0}^{t} d\tau\Theta_0(\tau)e^{-ikc\tau+i\vec{k}\cdot\overrightarrow{d_d'}(\tau)}\frac{d}{d\tau}\left[e^{i\vec{k}\cdot\overrightarrow{z_+'}(\tau)} - e^{i\vec{k}\cdot\overrightarrow{z_-'}(\tau)}\right]. \tag{C6}$$

Using the notations: $u = \Theta_0(\tau)e^{-ikc\tau+i\vec{k}\cdot\overrightarrow{d_d'}(\tau)} \Rightarrow \frac{du}{d\tau} = \delta(\tau)e^{-ikc\tau+i\vec{k}\cdot\overrightarrow{d_d'}(\tau)} - i\left[kc - \vec{k}\cdot\overrightarrow{\mathrm{v}_d}(\tau)\right]\Theta_0(\tau)e^{-ikc\tau+i\vec{k}\cdot\overrightarrow{d_d'}(\tau)}$, $\frac{dw}{d\tau} = \frac{d}{d\tau}\left[e^{i\vec{k}\cdot\overrightarrow{z_+'}(\tau)} - e^{i\vec{k}\cdot\overrightarrow{z_-'}(\tau)}\right] \Rightarrow w = e^{i\vec{k}\cdot\overrightarrow{z_+'}(\tau)} - e^{i\vec{k}\cdot\overrightarrow{z_-'}(\tau)}$, we integrate (C6) by parts and obtain

$$\vec{k}\cdot\overrightarrow{\mathcal{A}^*}(\vec{k},t) = -\frac{ec\mu_{\mathrm{v}}}{16\pi^3}\frac{1}{k}e^{-ikct+i\vec{k}\cdot\overrightarrow{d_d'}(t)}\left[e^{i\vec{k}\cdot\overrightarrow{z_+'}(t)} - e^{i\vec{k}\cdot\overrightarrow{z_-'}(t)}\right] + \frac{ec\mu_{\mathrm{v}}}{16\pi^3}\frac{1}{k}\int_{-\infty}^{t} d\tau\delta(\tau)e^{-ikc\tau+i\vec{k}\cdot\overrightarrow{d_d'}(\tau)}\left[e^{i\vec{k}\cdot\overrightarrow{z_+'}(\tau)} - e^{i\vec{k}\cdot\overrightarrow{z_-'}(\tau)}\right] - i\frac{ec\mu_{\mathrm{v}}}{16\pi^3}\int_{-\infty}^{t} d\tau\Theta_0(\tau)\left[c - \hat{k}\cdot\overrightarrow{\mathrm{v}_d}(\tau)\right]e^{-ikc\tau+i\vec{k}\cdot\overrightarrow{d_d'}(\tau)}\left[e^{i\vec{k}\cdot\overrightarrow{z_+'}(\tau)} - e^{i\vec{k}\cdot\overrightarrow{z_-'}(\tau)}\right].$$

or, after using the sifting property of the delta function in the second integral [together with $\overrightarrow{z_\pm'}(0) = 0$],

$$\vec{k}\cdot\overrightarrow{\mathcal{A}^*}(\vec{k},t) = -\frac{ec\mu_{\mathrm{v}}}{16\pi^3}\frac{1}{k}e^{-ikct+i\vec{k}\cdot\overrightarrow{d_d'}(t)}\left[e^{i\vec{k}\cdot\overrightarrow{z_+'}(t)} - e^{i\vec{k}\cdot\overrightarrow{z_-'}(t)}\right] - i\frac{e}{16\pi^3\varepsilon_{\mathrm{v}}}\int_{-\infty}^{t} d\tau\Theta_0(\tau)e^{-ikc\tau+i\vec{k}\cdot\overrightarrow{d_d'}(\tau)}\left[e^{i\vec{k}\cdot\overrightarrow{z_+'}(\tau)} - e^{i\vec{k}\cdot\overrightarrow{z_-'}(\tau)}\right] + i\frac{ec\mu_{\mathrm{v}}}{16\pi^3}\int_{-\infty}^{t} d\tau\Theta_0(\tau)\left[\hat{k}\cdot\overrightarrow{\mathrm{v}_d}(\tau)\right]e^{-ikc\tau+i\vec{k}\cdot\overrightarrow{d_d'}(\tau)}\left[e^{i\vec{k}\cdot\overrightarrow{z_+'}(\tau)} - e^{i\vec{k}\cdot\overrightarrow{z_-'}(\tau)}\right]. \tag{C7}$$

Using

$$\mathcal{F}^*(\vec{k},t) = -i\frac{ec\mu_{\mathrm{v}}}{16\pi^3}\frac{1}{k}\int_{-\infty}^{t} d\tau\Theta_0(\tau)e^{-ikc\tau+i\vec{k}\cdot\overrightarrow{d_d'}(\tau)}\left[e^{i\vec{k}\cdot\overrightarrow{z_+'}(\tau)} - e^{i\vec{k}\cdot\overrightarrow{z_-'}(\tau)}\right], \tag{18}$$

in Eqn. (C7), we obtain:

$$\vec{k}\cdot\overrightarrow{\mathcal{A}^*}(\vec{k},t) = -\frac{ec\mu_{\mathrm{v}}}{16\pi^3}\frac{1}{k}e^{-ikct+i\vec{k}\cdot\overrightarrow{d_d'}(t)}\left[e^{i\vec{k}\cdot\overrightarrow{z_+'}(t)} - e^{i\vec{k}\cdot\overrightarrow{z_-'}(t)}\right] + i\frac{ec\mu_{\mathrm{v}}}{16\pi^3}\int_{-\infty}^{t} d\tau\Theta_0(\tau)\left[\hat{k}\cdot\overrightarrow{\mathrm{v}_d}(\tau)\right]e^{-ikc\tau+i\vec{k}\cdot\overrightarrow{d_d'}(\tau)}\left[e^{i\vec{k}\cdot\overrightarrow{z_+'}(\tau)} - e^{i\vec{k}\cdot\overrightarrow{z_-'}(\tau)}\right] + \frac{k}{c}\mathcal{F}^*(\vec{k},t), \tag{C8}$$

or, equivalently,

$$\mathcal{F}^*(\vec{k},t) = c\hat{k}\cdot\overrightarrow{\mathcal{A}^*}(\vec{k},t) + \frac{e}{16\pi^3\varepsilon_{\mathrm{v}}}\frac{1}{k^2}e^{-ikct+i\vec{k}\cdot\overrightarrow{d_d'}(t)}\left[e^{i\vec{k}\cdot\overrightarrow{z_+'}(t)} - e^{i\vec{k}\cdot\overrightarrow{z_-'}(t)}\right] + c\hat{k}\cdot\vec{\mathcal{V}}^*(\vec{k},t), \tag{C9}$$

with

$$\vec{\mathcal{V}}^* = -i\frac{ec\mu_{\mathrm{v}}}{16\pi^3}\frac{1}{k}\int_{-\infty}^{t} d\tau\Theta_0(\tau)\overrightarrow{\mathrm{v}_d}(\tau)e^{-ikc\tau+i\vec{k}\cdot\overrightarrow{d_d'}(\tau)}\left[e^{i\vec{k}\cdot\overrightarrow{z_+'}(\tau)} - e^{i\vec{k}\cdot\overrightarrow{z_-'}(\tau)}\right]. \tag{C10}$$

Inserting Eqns. (C4) and (C9) into (16),

$$\phi(\vec{r},t) = \iiint_{-\infty}^{\infty} d^3k\left[\mathcal{F}(\vec{k},t)e^{i\vec{k}\cdot\vec{R}-ikct} + \mathcal{F}^*(\vec{k},t)e^{-i\vec{k}\cdot\vec{R}+ikct}\right], \tag{16}$$

we obtain

$$\phi(\vec{r},t) = \iiint_{-\infty}^{\infty} d^3k \Big\{ c\hat{k} \cdot \vec{\mathcal{A}}\big(\vec{k},t\big) e^{i\vec{k}\cdot\vec{R} - ikct} + \frac{e}{16\pi^3 \varepsilon_{\mathrm{v}}} \frac{1}{k^2} \Big[ e^{-i\vec{k}\cdot\overrightarrow{z'_+}(t) - i\vec{k}\cdot\overrightarrow{d'_d}(t)} - e^{-i\vec{k}\cdot\overrightarrow{z'_-}(t) - i\vec{k}\cdot\overrightarrow{d'_d}(t)} \Big] e^{i\vec{k}\cdot\vec{R}} + c\hat{k} \cdot \vec{\mathcal{V}}\big(\vec{k},t\big) e^{i\vec{k}\cdot\vec{R} - ikct} + c\hat{k} \cdot \overrightarrow{\mathcal{A}^*}\big(\vec{k},t\big) e^{-i\vec{k}\cdot\vec{R} + ikc} + \frac{e}{16\pi^3 \varepsilon_{\mathrm{v}}} \frac{1}{k^2} \Big[ e^{i\vec{k}\cdot\overrightarrow{z'_+}(t) + i\vec{k}\cdot\overrightarrow{d'_d}(t)} - e^{i\vec{k}\cdot\overrightarrow{z'_-}(t) + i\vec{k}\cdot\overrightarrow{d'_d}(t)} \Big] e^{-i\vec{k}\cdot\vec{R}} + c\hat{k} \cdot \vec{\mathcal{V}}^*\big(\vec{k},t\big) e^{-i\vec{k}\cdot\vec{R} + ikct} \Big\}, \tag{C11}$$

where we used the notations

$$\mathcal{V}\big(\vec{k},t\big) = i \frac{ec\mu_{\mathrm{v}}}{16\pi^3} \frac{1}{k} \int_{-\infty}^{t} d\tau \Theta_0(\tau) \overrightarrow{\mathrm{v}_d} e^{ikc\tau - \;\vec{}\;\cdot\overrightarrow{d'_d}(\tau)} \Big[ e^{-i\vec{k}\cdot\overrightarrow{z'_+}(\tau)} - e^{-i\vec{k}\cdot\overrightarrow{z'_-}(\tau)} \Big], \tag{24}$$

$$\vec{\mathcal{V}}^*\big(\vec{k},t\big) = -i \frac{ec\mu_{\mathrm{v}}}{16\pi^3} \frac{1}{k} \int_{-\infty}^{t} d\tau \Theta_0(\tau) \overrightarrow{\mathrm{v}_d}(\tau) e^{-ikc\tau + i\vec{k}\cdot\overrightarrow{d'_d}(\tau)} \Big[ e^{i\vec{k}\cdot\overrightarrow{z'_+}(\tau)} - e^{i\vec{k}\cdot\overrightarrow{z'_-}(\tau)} \Big]. \tag{26}$$

After grouping terms, the scalar potential becomes

$$\phi(\vec{r},t) = \iiint_{-\infty}^{\infty} d^3k \; c\hat{k} \cdot \Big[ \vec{\mathcal{A}}\big(\vec{k},t\big) e^{i\vec{k}\cdot\vec{R} - ik} \quad + \overrightarrow{\mathcal{A}^*}\big(\vec{k},t\big) e^{-i\vec{k}\cdot\vec{R} + ikct} \Big] + \frac{e}{16\pi^3 \varepsilon_{\mathrm{v}}} \iiint_{-\infty}^{\infty} d^3k \frac{1}{k^2} \Big[ e^{i\vec{k}\cdot\overrightarrow{R_+}(t)} + e^{-i\vec{k}\cdot\overrightarrow{R_+}(t)} - e^{i\vec{k}\cdot\overrightarrow{R_-}(t)} - e^{-i\vec{k}\cdot\overrightarrow{R_-}(t)} \Big] + \iiint_{-\infty}^{\infty} d^3k \; c\hat{k} \cdot \Big[ \vec{\mathcal{V}}\big(\vec{k},t\big) e^{i\vec{k}\cdot\vec{R} - ikct} + \vec{\mathcal{V}}^*\big(\vec{k},t\big) e^{-i\vec{k}\cdot\vec{R} + ikct} \Big]. \tag{C12}$$

But the exponentials in the second integrals may be grouped together into cosines, and the integrals may be solved to give, successively,

$$\frac{e}{16\pi^3 \varepsilon_{\mathrm{v}}} \iiint_{-\infty}^{\infty} d^3k \frac{1}{k^2} \Big[ e^{i\vec{k}\cdot\overrightarrow{R_+}(t)} + e^{-i\vec{k}\cdot\overrightarrow{R_+}(t)} - e^{i\vec{k}\cdot\overrightarrow{R_-}(t)} - e^{-i\vec{k}\cdot\overrightarrow{R_-}(t)} \Big] = \frac{e}{8\pi^3 \varepsilon_{\mathrm{v}}} \iiint_{-\infty}^{\infty} d^3k \frac{1}{k^2} \cos\big[\vec{k} \cdot \overrightarrow{R_+}(t)\big] - \frac{e}{8\pi^3 \varepsilon_{\mathrm{v}}} \iiint_{-\infty}^{\infty} d^3k \frac{1}{k^2} \cos\big[\vec{k} \cdot \overrightarrow{R_-}(t)\big] = \frac{2\pi e}{8\pi^3 \varepsilon_{\mathrm{v}}} \int_0^{\pi} d\theta \sin\theta \int_0^{\infty} dk \cos[kR_+(t)\cos\theta] - \frac{2\pi}{8\pi^3 \varepsilon_{\mathrm{v}}} \int_0^{\pi} d\theta \sin\theta \int_0^{\infty} dk \cos[kR_-(t)\cos\theta] = \frac{e}{2\pi^2 \varepsilon_{\mathrm{v}}} \int_0^{\infty} dk \frac{\sin kR_+(t)}{kR_+(t)} - \frac{e}{2\pi^2 \varepsilon_{\mathrm{v}}} \int_0^{\infty} dk \frac{\sin kR_-(t)}{kR_-(t)} = \frac{e}{2\pi^2 \varepsilon_{\mathrm{v}} R_+(t)} \int_0^{\infty} d[kR_+(t)] \frac{\sin kR_+(t)}{kR_+(t)} - \frac{e}{2\pi^2 \varepsilon_{\mathrm{v}} R_-(t)} \int_0^{\infty} d[kR_-(t)] \frac{\sin kR_-(t)}{kR_-(t)} = \frac{e}{4\pi\varepsilon_{\mathrm{v}} R_+(t)} - \frac{e}{4\pi\varepsilon_{\mathrm{v}} R_-(t)}, \tag{C13}$$

where we also used the fact that the integrals of the sinc functions are equal to $\pi/2$.

Inserting the last result in (C7-12), the k-space form of the potential separated into velocity-related and position-related terms may be written as

$$\phi(\vec{r},t) = \iiint_{-\infty}^{\infty} d^3k \; c\hat{k} \cdot \Big[ \vec{\mathcal{A}}\big(\vec{k},t\big) e^{i\vec{k}\cdot\vec{R} - ikct} + \overrightarrow{\mathcal{A}^*}\big(\vec{k},t\big) e^{-i\vec{k}\cdot\vec{R} + ikct} \Big] + \frac{e}{4\pi\varepsilon_{\mathrm{v}} R_+(t)} - \frac{e}{4\pi\varepsilon_{\mathrm{v}} R_-(t)} + \iiint_{-\infty}^{\infty} d^3k \; c\hat{k} \cdot \Big[ \vec{\mathcal{V}}\big(\vec{k},t\big) e^{i\vec{k}\cdot\vec{R} - ik} \quad + \vec{\mathcal{V}}^*\big(\vec{k},t\big) e^{-i\vec{k}\cdot\vec{R} + ikct} \Big]. \tag{C14}$$

This is Eqn. (27) in the main text.

**Appendix D: Derivation of the electric and magnetic fields given by equations (30) and (31)**

Since the scalar potential is taken as zero, the electric field originates from the vector potential given by

$$\vec{A}(\vec{r},t) = \iiint_{-\infty}^{\infty} d^3k \left[\vec{\mathcal{A}}\left(\vec{k},t\right)e^{i\vec{k}\cdot\vec{R}-ikc} + \overrightarrow{\mathcal{A}^{*}}\left(\vec{k},t\right)e^{-i\vec{k}\cdot\vec{R}+ikct}\right], \tag{19}$$

we have

$$-\frac{\partial}{\partial t}\vec{A} = ic\iiint_{-\infty}^{\infty} d^3k\, k\left[\vec{\mathcal{A}}\left(\vec{k},t\right)e^{i\vec{k}\cdot\vec{R}-ikc} - \overrightarrow{\mathcal{A}^{*}}\left(\vec{k},t\right)e^{-i\vec{k}\cdot\vec{R}+ikct}\right] - \iiint_{-\infty}^{\infty} d^3k \left[\frac{\partial}{\partial t}\vec{\mathcal{A}}\left(\vec{k},t\right)e^{i\vec{k}\cdot\vec{R}-ikc} + \frac{\partial}{\partial t}\overrightarrow{\mathcal{A}^{*}}\left(\vec{k},t\right)e^{-i\vec{k}\cdot\vec{R}+ikct}\right]. \tag{D1}$$

But, from equations

$$\vec{\mathcal{A}}\left(\vec{k},t\right) = i\frac{ec\mu_{\mathrm{v}}}{16\pi^3}\frac{1}{k}\int_{-\infty}^{t} d\tau\Theta_0(\tau)e^{ikc\tau-\vec{\ }\cdot\overrightarrow{d_d'}(\tau)}\left[\overrightarrow{\mathrm{v}_+}(\tau)e^{-i\vec{k}\cdot\overrightarrow{z_+'}(\tau)} - \overrightarrow{\mathrm{v}_-}(\tau)e^{-i\vec{k}\cdot\overrightarrow{z_-'}(\tau)}\right], \tag{20}$$

$$\overrightarrow{\mathcal{A}^{*}}\left(\vec{k},t\right) = -i\frac{ec\mu_{\mathrm{v}}}{16\pi^3}\frac{1}{k}\int_{-\infty}^{t} d\tau\Theta_0(\tau)e^{-ikc\tau+i\vec{k}\cdot\overrightarrow{d_d'}(\tau)}\left[\overrightarrow{\mathrm{v}_+}(\tau)e^{i\vec{k}\cdot\overrightarrow{z_+'}(\tau)} - \overrightarrow{\mathrm{v}_-}(\tau)e^{i\vec{k}\cdot\overrightarrow{z_-'}(\tau)}\right], \tag{21}$$

we have

$$\frac{\partial}{\partial t}\vec{\mathcal{A}}\left(\vec{k},t\right) = i\frac{ec\mu_{\mathrm{v}}}{16\pi^3}\frac{1}{k}\Theta_0(t)e^{ikct-i\vec{k}\cdot\overrightarrow{d_d'}(t)}\left[\overrightarrow{\mathrm{v}_+}(t)e^{-i\vec{k}\cdot\overrightarrow{z_+'}(t)} - \overrightarrow{\mathrm{v}_-}(t)e^{-i\vec{k}\cdot\overrightarrow{z_-'}(t)}\right], \tag{D2}$$

and

$$\frac{\partial}{\partial t}\overrightarrow{\mathcal{A}^{*}}\left(\vec{k},t\right) = -i\frac{ec\mu_{\mathrm{v}}}{16\pi^3}\frac{1}{k}\Theta_0(t)e^{-ikct+i\vec{k}\cdot\overrightarrow{d_d'}(t)}\left[\overrightarrow{\mathrm{v}_+}(t)e^{i\vec{k}\cdot\overrightarrow{z_+'}(t)} - \overrightarrow{\mathrm{v}_-}(t)e^{i\vec{k}\cdot\overrightarrow{z_-'}(t)}\right], \tag{D3}$$

which, inserted into (D1), give

$$-\frac{\partial}{\partial t}\vec{A} = i\iiint_{-\infty}^{\infty} d^3k\, kc\left[\vec{\mathcal{A}}\left(\vec{k},t\right)e^{i\vec{k}\cdot\vec{R}-ikc} - \overrightarrow{\mathcal{A}^{*}}\left(\vec{k},t\right)\right] - i\frac{ec\mu_{\mathrm{v}}}{16\pi^3}\iiint_{-\infty}^{\infty} d^3k\frac{1}{k}\Theta_0(t)\left[\overrightarrow{\mathrm{v}_+}(t)e^{-i\vec{k}\cdot\overrightarrow{R_+}(t)} - \overrightarrow{\mathrm{v}_-}(t)e^{-i\vec{k}\cdot\overrightarrow{R_-}(t)} - \overrightarrow{\mathrm{v}_+}(t)e^{i\vec{k}\cdot\overrightarrow{R_+}(t)} + \overrightarrow{\mathrm{v}_-}(t)e^{i\vec{k}\cdot\overrightarrow{R_-}(t)}\right]. \tag{D4}$$

Since the expression in the square bracket under the second integral may be simplified as

$$-2i\overrightarrow{\mathrm{v}_+}(t)\sin\left[\vec{k}\cdot\overrightarrow{R_+}(t)\right] + 2i\overrightarrow{\mathrm{v}_-}(t)\sin\left[\vec{k}\cdot\overrightarrow{R_-}(t)\right],$$

that integral becomes

$$\frac{ec\mu_{\mathrm{v}}}{8\pi^3}\iiint_{-\infty}^{\infty} d^3k\frac{1}{k}\Theta_0(t)\overrightarrow{\mathrm{v}_+}(t)\sin\left[\vec{k}\cdot\overrightarrow{R_+}(t)\right] - \frac{ec\mu_{\mathrm{v}}}{8\pi^3}\iiint_{-\infty}^{\infty} d^3k\frac{1}{k}\Theta_0(t)\overrightarrow{\mathrm{v}_-}(t)\sin\left[\vec{k}\cdot\overrightarrow{R_-}(t)\right].$$

If we were to switch to spherical coordinates in $\vec{k}$, by replacing $\vec{k}\cdot\overrightarrow{R_\pm}(\tau)$ by $kR_\pm(\tau)\cos\theta \equiv kR_\pm(\tau)\cos\theta$, the two integrals would become:

$$\iiint_{-\infty}^{\infty} d^3k\Theta_0(t)\overrightarrow{\mathrm{v}_\pm}(t)\sin\left[\vec{k}\cdot\overrightarrow{R_\pm}(\tau)\right] = \int_0^{2\pi} d\varphi\int_0^{\pi} d\theta\sin\theta\int_0^{\infty} dk\, k\Theta_0(t)\sin[kR_\pm(\tau)\cos\theta],$$

which, after using the substitutions $u = \cos\theta$ and $du = -\sin\theta\, d\theta$ and integrating, each becomes

$$2\pi\int_0^{\infty} dk\, k\int_1^{-1} du\sin[kR_\pm(\tau)u] = 0.$$

Thus, the second integral in Eqn. (D4) is zero, and we can write:

$$-\frac{\partial}{\partial t}\vec{A} = i\iiint_{-\infty}^{\infty} d^3k\, kc\left[\vec{\mathcal{A}}\left(\vec{k},t\right)e^{i\vec{k}\cdot\vec{R}-ikct} - \overrightarrow{\mathcal{A}^{*}}\left(\vec{k},t\right)e^{-i\vec{k}\cdot\vec{R}+ikct}\right]. \tag{D5}$$

The electric field is thus

$$\vec{E}(\vec{r},t) = i \iiint_{-\infty}^{\infty} d^3k\, kc \left[\vec{\mathcal{A}}\left(\vec{k},t\right)e^{i\vec{k}\cdot\vec{R}-ikct} - \overrightarrow{\mathcal{A}^{*}}\left(\vec{k},t\right)e^{-i\vec{k}\cdot\vec{R}+ikct}\right], \tag{D6}$$

which is Eqn. (30).

To derive the magnetic field, we apply $\vec{\nabla}\times$ to the vector potential,

$$\vec{A}(\vec{r},t) = \iiint_{-\infty}^{\infty} d^3k \left[\vec{\mathcal{A}}\left(\vec{k},t\right)e^{i\vec{k}\cdot\vec{R}-ikc} + \overrightarrow{\mathcal{A}^{*}}\left(\vec{k},t\right)e^{-i\vec{k}\cdot\vec{R}+ikct}\right], \tag{19}$$

and obtain

$$\vec{\nabla}\times\vec{A} = \iiint_{-\infty}^{\infty} d^3k\, \vec{\nabla}\times\left[\vec{\mathcal{A}}\left(\vec{k},t\right)e^{i\vec{k}\cdot\vec{R}-ikc} + \overrightarrow{\mathcal{A}^{*}}\left(\vec{k},t\right)e^{-i\vec{k}\cdot\vec{R}+ikct}\right] = \iiint_{-\infty}^{\infty} d^3k \left[\vec{\nabla}\times\vec{\mathcal{A}}\left(\vec{k},t\right)e^{i\vec{k}\cdot\vec{R}-ikct} + i\vec{\nabla}\left(\vec{k}\cdot\vec{R}\right)\times\vec{\mathcal{A}}\left(\vec{k},t\right)e^{i\vec{k}\cdot\vec{R}-ikc} + \vec{\nabla}\times\overrightarrow{\mathcal{A}^{*}}\left(\vec{k},t\right)e^{-i\vec{k}\cdot\vec{R}+ikc} - i\vec{\nabla}\left(\vec{k}\cdot\vec{R}\right)\times\overrightarrow{\mathcal{A}^{*}}\left(\vec{k},t\right)e^{-i\vec{k}\cdot\vec{R}+ikct}\right] = i \iiint_{-\infty}^{\infty} d^3k \left[\vec{\nabla}\left(\vec{k}\cdot\vec{R}\right)\times\vec{\mathcal{A}}\left(\vec{k},t\right)e^{i\vec{k}\cdot\vec{R}-ikct} - \vec{\nabla}\left(\vec{k}\cdot\vec{R}\right)\times\overrightarrow{\mathcal{A}^{*}}\left(\vec{k},t\right)e^{-i\vec{k}\cdot\vec{R}+ikct}\right]. \tag{D7}$$

From the relations

$$\vec{\mathcal{A}}\left(\vec{k},t\right) = i\frac{ec\mu_{\mathrm{v}}}{16\pi^3}\frac{1}{k}\int_{-\infty}^{t} d\tau\,\Theta_0(\tau)e^{ikc\tau-\ \vec{}\cdot\overrightarrow{d'_d}(\tau)}\left[\overrightarrow{\mathrm{v}_+}(\tau)e^{-i\vec{k}\cdot\overrightarrow{z'_+}(\tau)} - \overrightarrow{\mathrm{v}_-}(\tau)e^{-i\vec{k}\cdot\overrightarrow{z'_-}(\tau)}\right], \tag{20}$$

$$\overrightarrow{\mathcal{A}^{*}}\left(\vec{k},t\right) = -i\frac{ec\mu_{\mathrm{v}}}{16\pi^3}\frac{1}{k}\int_{-\infty}^{t} d\tau\,\Theta_0(\tau)e^{-ikc\tau+i\vec{k}\cdot\overrightarrow{d'_d}(\tau)}\left[\overrightarrow{\mathrm{v}_+}(\tau)e^{i\vec{k}\cdot\overrightarrow{z'_+}(\tau)} - \overrightarrow{\mathrm{v}_-}(\tau)e^{i\vec{k}\cdot\overrightarrow{z'_-}(\tau)}\right], \tag{21}$$

we obtain

$$\vec{\nabla}\times\vec{\mathcal{A}}\left(\vec{k},t\right) = \vec{\nabla}\times\overrightarrow{\mathcal{A}^{*}}\left(\vec{k},t\right) = 0.$$

In addition, using a well-known vector identity, we can write

$$\vec{\nabla}\left(\vec{k}\cdot\vec{R}\right) = \vec{k}\times\left(\vec{\nabla}\times\vec{R}\right) + \vec{R}\times\left(\vec{\nabla}\times\vec{k}\right) + \left(\vec{k}\cdot\vec{\nabla}\right)\vec{R} + \left(\vec{R}\cdot\vec{\nabla}\right)\vec{k} = 0 + 0 + \left(\vec{k}\cdot\vec{\nabla}\right)\vec{R} + 0, \tag{D8}$$

with

$$\left(\vec{k}\cdot\vec{\nabla}\right)\vec{R} = \vec{k}. \tag{D9}$$

Thus, Eqn. (D7) becomes

$$\vec{B}(\vec{r},t) = i \iiint_{-\infty}^{\infty} d^3k \left[\vec{k}\times\vec{\mathcal{A}}\left(\vec{k},t\right)\, e^{i\vec{k}\cdot\vec{R}-ikct} - \vec{k}\times\overrightarrow{\mathcal{A}^{*}}\left(\vec{k},t\right)\, e^{-i\vec{k}\cdot\vec{R}+ikc}\right], \tag{D10}$$

which is Eqn. (31).

**Appendix E: Proof that $q$ and $p$ given by Eqns. (45) & (46) obey Hamilton's relations (47) & (48)**

Replacing the upper integration limit in Eqns. (20) and (21) by $t_1$ – defined as the time at which emission of electromagnetic energy by the dipole is complete – and thus restricting the observation times to $t > t_1$, we get

$$\mathcal{A}(\vec{k}) = i\frac{e}{16\pi^3\varepsilon_\mathrm{V}}\frac{1}{\omega_k}\int_{-\infty}^{t_1} d\tau\Theta_0(\tau)e^{ikc\tau - i\vec{\ }\cdot\overrightarrow{d'_d}(\tau)}[\mathrm{v}_+(\tau) - \mathrm{v}_-(\tau)], \tag{E1}$$

$$\mathcal{A}^*(\vec{k}) = -i\frac{e}{16\pi^3\varepsilon_\mathrm{V}}\frac{1}{\omega_k}\int_{-\infty}^{t_1} d\tau\Theta_0(\tau)e^{-ikc\tau + i\vec{k}\cdot\overrightarrow{d'_d}(\tau)}[\mathrm{v}_+(\tau) - \mathrm{v}_-(\tau)]. \tag{E2}$$

Replacing $\mathcal{A}(\vec{k},t)$ and $\mathcal{A}^*(\vec{k},t)$ in

$$q(\vec{k},t) = \sqrt{\frac{25\quad{}^6\varepsilon_\mathrm{V}}{m\mathrm{V}}}\left[\alpha(\vec{k},t) + \alpha^*(\vec{k},t)\right], \tag{45}$$

$$p(\vec{k},t) = -i\omega_k\sqrt{\frac{256\pi^6\varepsilon_\mathrm{V}m}{\mathrm{V}}}\left[\alpha(\vec{k},t) - \alpha^*(\vec{k},t)\right], \tag{46}$$

by $\mathcal{A}(\vec{k})$ and $\mathcal{A}^*(\vec{k})$, we get

$$q(\vec{k},t) = \sqrt{\frac{256\quad{}^6\varepsilon_\mathrm{V}}{m\mathrm{V}}}\left[\mathcal{A}(\vec{k})e^{-ikct} + \mathcal{A}^*(\vec{k})e^{ikct}\right], \tag{E3}$$

$$p(\vec{k},t) = -i\omega_k\sqrt{\frac{256\pi^6\varepsilon_\mathrm{V}m}{\mathrm{V}}}\left[\mathcal{A}(\vec{k})e^{-ikc}\quad - \mathcal{A}^*(\vec{k})e^{ikct}\right]. \tag{E4}$$

Taking the derivatives with respect to *t* of Eqns. (E3) and (E4), we obtain immediately

$$\dot{q}(\vec{k},t) = \sqrt{\frac{256\quad{}^6\varepsilon_\mathrm{V}}{m\mathrm{V}}}\left[\dot{\mathcal{A}}(\vec{k})e^{-ikc}\quad + \dot{\mathcal{A}}^*(\vec{k})e^{ikct}\right] - i\omega_k\sqrt{\frac{25\quad{}^6\varepsilon_\mathrm{V}}{m\mathrm{V}}}\left[\mathcal{A}(\vec{k})e^{-ikc}\quad - \mathcal{A}^*(\vec{k})e^{ikct}\right], \tag{E5}$$

$$\dot{p}(\vec{k},t) = -\sqrt{\frac{256\pi^6\varepsilon_\mathrm{V}m}{\mathrm{V}}}\left\{i\omega_k\left[\dot{\mathcal{A}}(\vec{k})e^{-ikct} - \dot{\mathcal{A}}^*(\vec{k})e^{ikct}\right] - {\omega_k}^2\left[\mathcal{A}(\vec{k})e^{-ikc}\quad + \mathcal{A}^*(\vec{k})e^{ikct}\right]\right\}. \tag{E6}$$

which, since using (E1) and (E2) we get $\dot{\mathcal{A}}(k) = \dot{\mathcal{A}}^*(k) = 0$, become

$$\dot{q}(\vec{k},t) = -i\omega_k\sqrt{\frac{256\pi^6\varepsilon_\mathrm{V}}{m\mathrm{V}}}\left[\mathcal{A}(\vec{k})e^{-ikct} - \mathcal{A}^*(\vec{k})e^{ikct}\right], \tag{E7}$$

$$\dot{p}(\vec{k},t) = -{\omega_k}^2\sqrt{\frac{256\pi^6\varepsilon_\mathrm{V}m}{\mathrm{V}}}\left[\mathcal{A}(\vec{k})e^{-ikct} + \mathcal{A}^*(\vec{k})e^{ikct}\right]. \tag{E8}$$

Comparison of Eqns. (E7) and (E8) to Eqns. (E3) and (E4), gives the sought after result:

$$\dot{q}(\vec{k},t) = p(t)\frac{\sqrt{\frac{256\pi^6\varepsilon_\mathrm{V}}{m\mathrm{V}}}}{\sqrt{\frac{256\pi^6\varepsilon_\mathrm{V}m}{\mathrm{V}}}} = \frac{1}{m}p(\vec{k},t), \tag{E9}$$

$$\dot{p}(\vec{k},t) = -\frac{{\omega_k}^2\sqrt{\frac{25\quad{}^6\varepsilon_\mathrm{V}m}{\mathrm{V}}}}{\sqrt{\frac{256\pi^6\varepsilon_\mathrm{V}}{m\mathrm{V}}}} = -m{\omega_k}^2q(\vec{k},t). \tag{E10}$$

**Appendix F: Commutation relation for the creation and annihilation operators**

Written in terms of the quantum mechanics operators,

$$\mathcal{A}(\vec{k}) \rightarrow \left(\frac{\hbar \mathrm{V}}{128\pi^{6}\varepsilon_{\mathrm{v}}\omega_{k}}\right)^{1/2} \boldsymbol{a}(\vec{k}), \tag{49}$$

$$\mathcal{A}^{*}(\vec{k}) \rightarrow \left(\frac{\hbar \mathrm{V}}{128\pi^{6}\varepsilon_{\mathrm{v}}\omega_{k}}\right)^{1/2} \boldsymbol{a}^{\dagger}(\vec{k}), \tag{50}$$

equations

$$\mathcal{A}(\vec{k})e^{-ik} \quad = \frac{\mathrm{V}^{1/2}}{(256\pi^{6}\varepsilon_{\mathrm{v}}m)^{1/2}\omega_{k}}[m\omega_{k}q + ip], \tag{42}$$

$$\mathcal{A}^{*}(\vec{k})e^{ikct} = \frac{\mathrm{V}^{1/2}}{(256\pi^{6}\varepsilon_{\mathrm{v}}m)^{1/2}\omega_{k}}[m\omega_{k}q - ip], \tag{43}$$

become

$$\boldsymbol{a}(\vec{k}) = \frac{1}{(2\hbar\omega_{k}m)^{1/2}}(m\omega_{k}\boldsymbol{q} + i\boldsymbol{p})e^{-ikct}, \tag{F1}$$

$$\boldsymbol{a}^{\dagger}(\vec{k}) = \frac{1}{(2\hbar\omega_{k}m)^{1/2}}(m\omega_{k}\boldsymbol{q} - i\boldsymbol{p})e^{ikct}. \tag{F2}$$

Using these two equations, we may write successively

$$\boldsymbol{a}\boldsymbol{a}^{\dagger} = \frac{1}{2\hbar\omega_{k}m}[m^{2}\omega_{k}{}^{2}\boldsymbol{q}\boldsymbol{q} - im\omega_{k}(\boldsymbol{q}\boldsymbol{p} - \boldsymbol{p}\boldsymbol{q}) + \boldsymbol{p}\boldsymbol{p}], \tag{F3}$$

$$\boldsymbol{a}^{\dagger}\boldsymbol{a} = \frac{1}{2m\hbar\omega_{k}}[m^{2}\omega_{k}{}^{2}\boldsymbol{q}\boldsymbol{q} + im\omega_{k}(\boldsymbol{q}\boldsymbol{p} - \boldsymbol{p}\boldsymbol{q}) + \boldsymbol{p}\boldsymbol{p}]. \tag{F4}$$

Using the last two expressions and the canonical commutator [1,2]

$$[\boldsymbol{q}, \boldsymbol{p}] = i\hbar, \tag{F5}$$

we obtain

$$[\boldsymbol{a}, \boldsymbol{a}^{\dagger}] \equiv \boldsymbol{a}\boldsymbol{a}^{\dagger} - \boldsymbol{a}^{\dagger}\boldsymbol{a} = -\frac{i}{2m\hbar\omega_{k}}m\omega_{k}(\boldsymbol{q}\boldsymbol{p} - \boldsymbol{p}\boldsymbol{q}) - \frac{i}{2m\hbar\omega_{k}}m\omega_{k}(\boldsymbol{q}\boldsymbol{p} - \boldsymbol{p}\boldsymbol{q}) = -\frac{i}{2\hbar}i\hbar - \frac{i}{2\hbar}i\hbar = 1. \tag{F6}$$

### Appendix G: Emission stimulated by a photon from a single-mode laser

The combined stimulating (subscript "s") fields and the fields emitted by the already excited dipole (subscript "d"),

$$\vec{\boldsymbol{E}}_t = \vec{\boldsymbol{E}}_s + \vec{\boldsymbol{E}}_d, \tag{G1}$$

$$\vec{\boldsymbol{B}}_t = \vec{\boldsymbol{B}}_s + \vec{\boldsymbol{B}}_d, \tag{G2}$$

carry electromagnetic energy integrated over the entire space centered around the dipole given by

$$\frac{\varepsilon_{\mathrm{v}}}{2} \iiint_{-\infty}^{\infty} d^3R\, \vec{\boldsymbol{E}}_t \cdot \vec{\boldsymbol{E}}_t + \frac{1}{2\mu_{\mathrm{v}}} \iiint_{-\infty}^{\infty} d^3R\, \vec{\boldsymbol{B}}_t \cdot \vec{\boldsymbol{B}}_t = \frac{\varepsilon_{\mathrm{v}}}{2} \iiint_{-\infty}^{\infty} d^3R \left(\vec{\boldsymbol{E}}_s \cdot \vec{\boldsymbol{E}}_s + \vec{\boldsymbol{E}}_d \cdot \vec{\boldsymbol{E}}_d + 2\vec{\boldsymbol{E}}_s \cdot \vec{\boldsymbol{E}}_d\right) + \frac{1}{2\mu_{\mathrm{v}}} \iiint_{-\infty}^{\infty} d^3R \left(\vec{\boldsymbol{B}}_s \cdot \vec{\boldsymbol{B}}_s + \vec{\boldsymbol{B}}_d \cdot \vec{\boldsymbol{B}}_d + 2\vec{\boldsymbol{B}}_s \cdot \vec{\boldsymbol{B}}_d\right). \tag{G3}$$

We use the expressions

$$\vec{\boldsymbol{E}}_d(\vec{r},t) = -i\left(\frac{\hbar \mathrm{V}}{128\pi^6 \varepsilon_{\mathrm{v}}}\right)^{1/2} \iiint_{-\infty}^{\infty} d^3k \left[\hat{k} \times \left(\hat{k} \times \hat{z}\right)\right] \omega_k{}^{1/2} \left[\boldsymbol{a}\left(\vec{k}\right) e^{i\vec{k}\cdot\vec{R}-ikc} - \boldsymbol{a}^{\dagger}\left(\vec{k}\right) e^{-i\vec{k}\cdot\vec{R}+ikct}\right], \tag{82}$$

$$\vec{\boldsymbol{B}}_d(\vec{r},t) = i\left(\frac{\hbar \mathrm{V}}{128\pi^6 \varepsilon_{\mathrm{v}}}\right)^{1/2} \iiint_{-\infty}^{\infty} d^3k\, k\left(\hat{k} \times \hat{z}\right) \omega_k{}^{-1/2} \left[\boldsymbol{a}\left(\vec{k}\right) e^{i\vec{k}\cdot\vec{R}-ikc} - \boldsymbol{a}^{\dagger}\left(\vec{k}\right) e^{-i\vec{k}\cdot\vec{R}+ik}\right], \tag{83}$$

for the dipole field, and

$$\vec{\boldsymbol{E}}_s(\vec{r},t) = i\left(\frac{\hbar}{2\varepsilon_{\mathrm{v}} \mathrm{V}}\right)^{1/2} \hat{z}\, \omega_s{}^{1/2} \left[\boldsymbol{a}\left(\vec{k}_s\right) e^{i\vec{k}_s\cdot\vec{R}-ik_sct} - \boldsymbol{a}^{\dagger}\left(\vec{k}_s\right) e^{-i\vec{k}_s\cdot\vec{R}+ik_sct}\right], \tag{98}$$

$$\vec{\boldsymbol{B}}_s(\vec{r},t) = i\left(\frac{\hbar}{2\varepsilon_{\mathrm{v}} \mathrm{V}}\right)^{1/2} k_s\left(\hat{k}_s \times \hat{z}\right) \omega_s{}^{-1/2} \left[\boldsymbol{a}\left(\vec{k}_s\right) e^{i\vec{k}_s\cdot\vec{R}-ik_sct} - \boldsymbol{a}^{\dagger}\left(\vec{k}_s\right) e^{-i\vec{k}_s\cdot\vec{R}+ik_sct}\right], \tag{99}$$

for the stimulating field. Note that $\vec{\boldsymbol{E}}_s$ commutes with $\vec{\boldsymbol{E}}_d$ and $\vec{\boldsymbol{B}}_s$ commutes with $\vec{\boldsymbol{B}}_d$. In addition, we assume that there is no phase difference between the stimulating and dipole radiator fields.

The third terms in each of the energy integrals, which are associated with the interference between the stimulating and radiated fields lead to

$$\boldsymbol{H}_{SE} = \varepsilon_{\mathrm{v}} \iiint_{-\infty}^{\infty} d^3R \left(\vec{\boldsymbol{E}}_s \cdot \vec{\boldsymbol{E}}_d\right) + \frac{1}{\mu_{\mathrm{v}}} \iiint_{-\infty}^{\infty} d^3R \left(\vec{\boldsymbol{B}}_s \cdot \vec{\boldsymbol{B}}_d\right) = \frac{\hbar}{16\pi^3} \iiint_{-\infty}^{\infty} d^3k\, \omega_s{}^{1/2} \omega_k{}^{1/2}\, \hat{z} \cdot \left[\hat{k} \times \left(\hat{k} \times \hat{z}\right)\right] \iiint_{-\infty}^{\infty} d^3R \left[\boldsymbol{a}\left(\vec{k}_s\right) e^{i\vec{k}_s\cdot\vec{R}-ik_sct} - \boldsymbol{a}^{\dagger}\left(\vec{k}_s\right) e^{-i\vec{k}_s\cdot\vec{R}+ik_sct}\right] \left[\boldsymbol{a}\left(\vec{k}\right) e^{i\vec{k}\cdot\vec{R}-ikct} - \boldsymbol{a}^{\dagger}\left(\vec{k}\right) e^{-i\vec{k}\cdot\vec{R}+ikct}\right] - \frac{\hbar}{16\pi^3 \varepsilon_{\mathrm{v}}} \frac{1}{\mu_{\mathrm{v}}} \iiint_{-\infty}^{\infty} d^3k\, k_s k\, \omega_s{}^{-\frac{1}{2}} \omega_k{}^{-\frac{1}{2}} \left(\hat{k}_s \times \hat{z}\right) \cdot \left(\hat{k} \times \hat{z}\right) \iiint_{-\infty}^{\infty} d^3R \left[\boldsymbol{a}\left(\vec{k}_s\right) e^{i\vec{k}_s\cdot\vec{R}-ik_sct} - \boldsymbol{a}^{\dagger}\left(\vec{k}_s\right) e^{-i\vec{k}_s\cdot\vec{R}+ik_sct}\right] \left[\boldsymbol{a}\left(\vec{k}\right) e^{i\vec{k}\cdot\vec{R}-ikct} - \boldsymbol{a}^{\dagger}\left(\vec{k}\right) e^{-i\vec{k}\cdot\vec{R}+ikct}\right] = \frac{\hbar}{16\pi^3} \iiint_{-\infty}^{\infty} d^3k\, \omega_s{}^{\frac{1}{2}} \omega_k{}^{\frac{1}{2}}\, \hat{z} \cdot \left[\hat{k} \times \left(\hat{k} \times \hat{z}\right)\right] \iiint_{-\infty}^{\infty} d^3R \left[\boldsymbol{a}\left(\vec{k}_s\right) \boldsymbol{a}\left(\vec{k}\right) e^{i\vec{k}\cdot\vec{R}-ikct+i\vec{k}_s\cdot\vec{R}-ik_sct} - \boldsymbol{a}\left(\vec{k}_s\right) \boldsymbol{a}^{\dagger}\left(\vec{k}\right) e^{-i\vec{k}\cdot\vec{R}+ikct+i\vec{k}_s\cdot\vec{R}-ik_sct} - \boldsymbol{a}^{\dagger}\left(\vec{k}_s\right) \boldsymbol{a}\left(\vec{k}\right) e^{i\vec{k}\cdot\vec{R}-ikct-i\vec{k}_s\cdot\vec{R}+ik_sct} + \boldsymbol{a}^{\dagger}\left(\vec{k}_s\right) \boldsymbol{a}^{\dagger}\left(\vec{k}\right) e^{-i\vec{k}\cdot\vec{R}+ikct-i\vec{k}_s\cdot\vec{R}+ik_sct}\right] -$$

$$\frac{\hbar c^2}{16\pi^3} \iiint_{-\infty}^{\infty} d^3k\ k_s k \omega_s{}^{-1/2} \omega_k{}^{-1/2} \left(\hat{k}_s \times \hat{z}\right) \cdot \left(\hat{k} \times \hat{z}\right) \iiint_{-\infty}^{\infty} d^3R \left[ \boldsymbol{a}\left(\vec{k}_s\right) \boldsymbol{a}\left(\vec{k}\right) e^{i\vec{k}\cdot\vec{R}-ikct+i\vec{k}_s\cdot\vec{R}-ik_sct} - \right.$$

$$\boldsymbol{a}\left(\vec{k}_s\right) \boldsymbol{a}^\dagger\left(\vec{k}\right) e^{-i\vec{k}\cdot\vec{R}+ikct+i\vec{k}_s\cdot\vec{R}-ik_sct} - \boldsymbol{a}^\dagger\left(\vec{k}_s\right) \boldsymbol{a}\left(\vec{k}\right) e^{i\vec{k}\cdot\vec{R}-ikct-i\vec{k}_s\cdot\vec{R}+ik_sct} +$$

$$\left. \boldsymbol{a}^\dagger\left(\vec{k}_s\right) \boldsymbol{a}^\dagger\left(\vec{k}\right) e^{-i\vec{k}\cdot\vec{R}+ikct-i\vec{k}_s\cdot\vec{R}+ik_sct} \right]. \qquad \text{(G4)}$$

Recognizing the integrals over space as the Dirac delta functions of the type defined by Eqns. (37) and (38), and using the sifting property of the delta function, we obtain successively:

$$\boldsymbol{H}_{SE} = \frac{1}{2}\hbar \iiint_{-\infty}^{\infty} d^3k\ \omega_s{}^{1/2} \omega_k{}^{1/2}\ \hat{z} \cdot \left[\hat{k} \times \left(\hat{k} \times \hat{z}\right)\right] \left[ \boldsymbol{a}\left(\vec{k}_s\right) \boldsymbol{a}\left(\vec{k}\right) e^{-ikct-i\omega_s t} \delta^3\left(\vec{k}+\vec{k}_s\right) - \right.$$

$$\boldsymbol{a}\left(\vec{k}_s\right) \boldsymbol{a}^\dagger\left(\vec{k}\right) e^{ikct- \quad {}_s t} \delta^3\left(\vec{k}-\vec{k}_s\right) - \boldsymbol{a}^\dagger\left(\vec{k}_s\right) \boldsymbol{a}\left(\vec{k}\right) e^{-ikct+i\omega_s t} \delta^3\left(\vec{k}-\vec{k}_s\right) +$$

$$\left. \boldsymbol{a}^\dagger\left(\vec{k}_s\right) \boldsymbol{a}^\dagger\left(\vec{k}\right) e^{ikct+i\omega_s t} \delta^3\left(\vec{k}+\vec{k}_s\right) \right] - \frac{1}{2}\hbar \iiint_{-\infty}^{\infty} d^3k\ \omega_s{}^{1/2} \omega_k{}^{1/2} \left(\hat{k}_s \times \hat{z}\right) \cdot \left(\hat{k} \times \right.$$

$$\left. \hat{z}\right) \left[ \boldsymbol{a}\left(\vec{k}_s\right) \boldsymbol{a}\left(\vec{k}\right) e^{-ikct-i\omega_s t} \delta^3\left(\vec{k}+\vec{k}_i\right) - \boldsymbol{a}\left(\vec{k}_s\right) \boldsymbol{a}^\dagger\left(\vec{k}\right) e^{ikct-i\omega_s t} \delta^3\left(\vec{k}-\vec{k}_s\right) - \right.$$

$$\left. \boldsymbol{a}^\dagger\left(\vec{k}_s\right) \boldsymbol{a}\left(\vec{k}\right) e^{-ikct+i\omega_s t} \delta^3\left(\vec{k}-\vec{k}_s\right) + \boldsymbol{a}^\dagger\left(\vec{k}_s\right) \boldsymbol{a}^\dagger\left(\vec{k}\right) e^{ikct+i\omega_s t} \delta^3\left(\vec{k}+\vec{k}_s\right) \right] = \frac{1}{2}\hbar\omega_s\ \hat{z} \cdot \left[\hat{k}_s \times \right.$$

$$\left. \left(\hat{k}_s \times \hat{z}\right)\right] \left[ \boldsymbol{a}\left(\vec{k}_s\right) \boldsymbol{a}\left(-\vec{k}_s\right) - \boldsymbol{a}\left(\vec{k}_s\right) \boldsymbol{a}^\dagger\left(\vec{k}_s\right) - \boldsymbol{a}^\dagger\left(\vec{k}_s\right) \boldsymbol{a}\left(\vec{k}_s\right) + \boldsymbol{a}^\dagger\left(\vec{k}_s\right) \boldsymbol{a}^\dagger\left(-\vec{k}_s\right) \right] - \frac{1}{2}\hbar\omega_s \left(\hat{k}_s \times \hat{z}\right) \cdot$$

$$\left(\hat{k}_s \times \hat{z}\right) \left[ -\boldsymbol{a}\left(\vec{k}_s\right) \boldsymbol{a}\left(-\vec{k}_s\right) - \boldsymbol{a}\left(\vec{k}_s\right) \boldsymbol{a}^\dagger\left(\vec{k}_s\right) - \boldsymbol{a}^\dagger\left(\vec{k}_s\right) \boldsymbol{a}\left(\vec{k}_s\right) - \boldsymbol{a}^\dagger\left(\vec{k}_s\right) \boldsymbol{a}^\dagger\left(-\vec{k}_s\right) \right]. \qquad \text{(G5)}$$

Using the vector identities $\vec{a} \times \left(\vec{b} \times \vec{c}\right) = (\vec{a} \cdot \vec{c})\vec{b} - \left(\vec{a} \cdot \vec{b}\right)\vec{c}$ and $\left(\vec{a} \times \vec{b}\right) \cdot \left(\vec{c} \times \vec{d}\right) = (\vec{a} \cdot \vec{c})\left(\vec{b} \cdot \vec{d}\right) - \left(\vec{a} \cdot \vec{d}\right)\left(\vec{b} \cdot \vec{c}\right)$, we get

$$\boldsymbol{H}_{SE} = \frac{1}{2}\hbar\omega_s \left[\left(\hat{z} \cdot \hat{k}_s\right)^2 - 1\right] \left[ \boldsymbol{a}\left(\vec{k}_s\right) \boldsymbol{a}\left(-\vec{k}_s\right) - \boldsymbol{a}\left(\vec{k}_s\right) \boldsymbol{a}^\dagger\left(\vec{k}_s\right) - \boldsymbol{a}^\dagger\left(\vec{k}_s\right) \boldsymbol{a}\left(\vec{k}_s\right) + \boldsymbol{a}^\dagger\left(\vec{k}_s\right) \boldsymbol{a}^\dagger\left(-\vec{k}_s\right) \right] -$$

$$\frac{1}{2}\hbar\omega_s \left[1 - \left(\hat{z} \cdot \hat{k}_s\right)^2\right] \left[ -\boldsymbol{a}\left(\vec{k}_s\right) \boldsymbol{a}\left(-\vec{k}_s\right) - \boldsymbol{a}\left(\vec{k}_s\right) \boldsymbol{a}^\dagger\left(\vec{k}_s\right) - \boldsymbol{a}^\dagger\left(\vec{k}_s\right) \boldsymbol{a}\left(\vec{k}_s\right) - \boldsymbol{a}^\dagger\left(\vec{k}_s\right) \boldsymbol{a}^\dagger\left(-\vec{k}_s\right) \right] =$$

$$\hbar\omega_s \sin^2 \varkappa_{\vec{k}S} \left[ \boldsymbol{a}\left(\vec{k}_s\right) \boldsymbol{a}^\dagger\left(\vec{k}_s\right) + \boldsymbol{a}^\dagger\left(\vec{k}_s\right) \boldsymbol{a}\left(\vec{k}_s\right) \right] = 2\hbar\omega_S \sin^2 \varkappa_{\vec{k}S} \left[ \boldsymbol{a}^\dagger\left(\vec{k}_s\right) \boldsymbol{a}\left(\vec{k}_s\right) + \frac{1}{2} \right]. \qquad \text{(G6)}$$